\documentclass[%
 reprint,
 amsmath,amssymb,
 aps,
]{revtex4-2}

\usepackage{graphicx}
\usepackage{bm}
\usepackage{dcolumn}
\usepackage{float}
\usepackage{caption}
\usepackage{subfigure}
\usepackage{amsmath}
\usepackage{mathrsfs}
\usepackage{amssymb}
\usepackage{hyperref}
\usepackage{xcolor}

\begin{document}

\preprint{}

\title{Non-electrostatic interactions as random fields in charged liquids}
\author{Li Wan}
\email{lwan@wzu.edu.cn}
\affiliation{Department of Physics, Wenzhou University, Wenzhou 325035, P. R. China}%


\begin{abstract}
We derive an equation capable of treating both the electrostatic and non-electrostatic interactions in the charged liquids. The equation derived is a complex version of the Poisson-Boltzmann equation, in which the non-electrostatic interactions are transformed to complex random fields by the field theory. Thus, the non-electrostatic interactions of the ions in the charged liquids can be simulated easily by generating random numbers according to the random fields. To solve the equation, the finite element method has been applied. The non-electrostatic interactions treated in the equation are general. In this study, we take the steric effect of ions as an example. Results show that the steric effect can be clearly caught by the equation.
\end{abstract}

\maketitle


\section{introduction}
The non-electrostatic interactions(NEIs) between charged objects in the charged liquids is of great importance in biological systems and industrial applications ~\cite{Israelachvili,Honig,Langmuir,Manciu,Parsons,Wennerstrom}. The NEIs include Lennard-Jones interaction, van der Waals interaction, steric effect and hydration forces between the charged objects~\cite{Israelachvili,Fedorov,Klaassen,Parsegian}. The steric effect is referred to the charged objects in the liquids excluding each other due to their finite size. The competition between the NEIs and the electrostatic interaction (EI) for the charged objects controls many processes, such as the stabilization of charged objects, electro-chemistry. In the charged liquids, it is well known that the charged objects are screened by ions through the EI, which is the Debye screening~\cite{Debye}. The NEIs can make the Debye screening anomalous~\cite{Gebbie,Smith,Lee,Coupette,Bazant}. In order to model a real charged liquid, the NEIs must be considered. \\

Extensive theoretical works have been carried out on this issue. In these theories, the computational simulations such as the molecular dynamics simulations or the Monte Carlo simulations for the charged liquids have been widely used. It is very convenient for the computational simulations to incorporate the NEIs into the interaction potential between ions~\cite{Blum,Jeanmairet,Kondrat,Zeman,Guldbrand,Joensson}. However, the computational simulations are computationally intensive.\\

Poisson-Boltzmann(PB) equation is a powerful tool and has been widely used to study the EI in the charged liquids. The PB equation is a mean field theory assuming that the ions in the liquids are point-like and the correlations between the ions are neglected. In the liquids with dilute ions, the PB equation can be linearized and analytical solutions can be obtained. When the ionic strength is increased in the liquids, the steric effect of the ions deviates the ionic profile from the results obtained by the PB equation~\cite{Loubet,Bazant,Maggs,Lue,Borukhov}. In order to catch the steric effects, the PB equation has to be modified. Various methods have been proposed to modify the PB equation~\cite{Loubet,Bazant,Maggs,Lue,Borukhov}. One of the methods is to use the lattice gas model to simulate the charged liquids~\cite{Borukhov,Bazant,Maggs}. In the method, the steric effect can be clearly involved in the entropy contribution to the free energy of the system. The lattice gas model is very convenient for the study of the steric effect, but it is not the real configuration of the liquids. The lattice gas model is also not general if the NEIs other than the steric effect are considered. To treat the NEIs generally, it has been proposed to separate the whole system of a charged liquid into two subsystems~\cite{Lue}. One subsystem is with the NEIs only and the other subsystem is with the EI alone. The subsystem with the NEIs is set as a reference system and the subsystem of the EI is considered as the perturbation. For the reference system, a precise formula is required to describe the equation of state of the ions. For various liquids, the formula varies and depends on the states of the liquids.\\

In the aspect of the EI, it has been observed that the ion-ion correlations through the Coulomb interaction bring many novel phenomena such as the like-charge attraction, aggregation of polymers and condensation of DNA~\cite{1Lau,1Linse,1Moreira,1Pelta,1Takahashi,1Yoshikawa}. To catch the ion-ion correlations, the field theory is a very versatile tool~\cite{2Naji,2Netz1,2Netz2,2Netz3,2Podgornik1,2Podgornik2}. The field theory expresses the partition function of the charged liquids by auxiliary  fields. The saddle point solution to the partition function is exactly the PB equation. The ion-ion correlation can be incorporated into the PB equation by the loop expansion or variational approach of the partition functions~\cite{2Buyukdagli1,2Buyukdagli2,2Buyukdagli3,2Moreira,2Naji1}. Especially, the variational approach combined with the cut-off of the Fourier transformation of the grand potential can be used to study both the effect of the ion-ion correlation and the steric effect~\cite{Loubet}. A different way to study the ion-ion correlation is to introduce a random field in a modified PB equation~\cite{Wan1}. It is easy to generate random numbers for the ion-ion correlations. Till now, in such a modified PB equation with the random field, the NEIs are still missing~\cite{Wan1}. \\

In this study, we start from the partition function of the charged liquids, and use the field theory to cast the NEIs and the ion fluctuations of the EI to two random fields. Then, we derive a complex differential equation with the two random fields involved, which is the complex version of the PB equation. In the derivation, only the functions of the pairwise potentials between the ions are needed for the NEIs. In the complex equation obtained, both the NEIs and the EI can be treated, including the ion-ion correlations. The equation is very simple and general for all the NEIs. To solve the complex differential equation, the finite element method is suggested.

\section{theory}
We consider a charged liquid consisting of only two ion species with opposite charges to demonstrate our theory. For the positive ions, the total ion number is $N_+$ and the charge value is $z_+$. And for the negative ions, they are $N_-$ and $z_-$ respectively. Note that $z_+$ is a positive value and $z_-$ is a negative value. The position vector of the $j$-th ion is denoted by $\vec{r}_j$. We define the number density for the positive ions by $c_{+}(\vec{r})=\sum_{j=1}^{N_{+}}\delta(\vec{r}-\vec{r}_j)$  and for the negative ions by $c_{-}(\vec{r})=\sum_{j=1}^{N_{-}}\delta(\vec{r}-\vec{r}_j)$. To save symbols, the notation $\vec{r}_j$ in the definition of $c_{+}(\vec{r})$ is for the positive ions only and $\vec{r}_j$ in $c_{-}(\vec{r})$ is for the negative ions only. Then, the number density of the total ions is $p(\vec{r})=c_+(\vec{r})+c_-(\vec{r})$. The net charge density in the system could be obtained as $e\rho(\vec{r})=e\sigma+e[z_+ c_++z_- c_-]$ with $e$ the elementary positive charge. Here, $e \sigma$ is the charge density of the external charges in the solid boundaries. \\

The Coulomb interaction between two ions located at  $\vec{r}$ and $\vec{r}'$ respectively is denoted by $C(\vec{r},\vec{r}')$, which is the Green function satisfying the equation $-\nabla_r\cdot[\epsilon \nabla_r C(\vec{r},\vec{r}')]=\delta(\vec{r}-\vec{r}')$. The operator $\nabla_r$ is applied in the real space of $\vec{r}$. $\epsilon$ is the dielectric constant of the liquid. It can be solved out that $C^{-1}=\nabla_r\cdot[\epsilon \nabla_{r'}\delta(\vec{r}-\vec{r}')]$, which can be checked through $\int d\vec{r}'' C^{-1}(\vec{r},\vec{r}'')C(\vec{r}'',\vec{r}')=\delta(\vec{r}-\vec{r}')$. To simplify our present study, the NEIs in the charged liquid are pairwise. All the pairwise NEIs between two given ions located at  $\vec{r}$ and $\vec{r}'$ respectively can be added up. The sum of the NEIs is denoted by $D(\vec{r},\vec{r}')$. Note that $D(\vec{r},\vec{r}')$ is translational invariant, and dependent on the displacement $\vec{x}=\vec{r}-\vec{r}'$.\\

Before we go further, we need define two pairs of Fourier transformation(FT). One pair of the FT is $g(G)=\int p(\vec{r})e^{-i G\vec{r}}d\vec{r}$ and $p(\vec{r})=\frac{1}{(2\pi)^3}\int g(G)e^{i G\vec{r}}dG$. The vector $G$ is defined in the reciprocal space and $i$ is the imaginary unit. Thus, we obtain $g(G)=g_++g_-$ with $g_{\pm}=\sum_{j=1}^{N_{\pm}}e^{-iG\vec{r}_j}$. The other pair of the FT is $T'(G)=\int D(\vec{x}) e^{-iG\vec{x}}d\vec{x}$ and  $D(\vec{x})=\frac{1}{(2\pi)^3}\int T'(G) e^{iG\vec{x}}dG$. We rewrite $T'(G)=(2\pi)^3\lambda T$ with $T=|T'(G)|/(2\pi)^3$ and $\lambda$ representing the sign of $T'(G)$. It is $\lambda=+1$ if $T'(G)$ is positive, and $\lambda=-1$ if $T'(G)$ is negative. Note that $\lambda$ and $T$ both are functional of the vector $G$. For the derivation, the following Hubbard-Stratonovich transformation(HST) will be used $e^{-\lambda a^2/2}=\frac{1}{Z_{y}}\int dy~ e^{-y^2/2+\sqrt{-\lambda}y a}$ with $Z_{y}=\int dy~ e^{-y^2/2}$ and $\lambda$ taking the value of $+1$ or $-1$.\\

\subsection{partition function}
The canonical partition function of the system is
\begin{align}
\label{Q}
Q=\frac{1}{N_+!N_-!\lambda_+^{3N_+}\lambda_-^{3N_-}}\int \prod_{s=1}^{N_+}d\vec{r}_s\prod_{j=1}^{N_-}d\vec{r}_je^{A}
\end{align}
with $A=-\beta H+e\int d\vec{r}h(\vec{r})\rho(\vec{r})$ .
Here, $\lambda_+$ and $\lambda_-$ are the de Broglie wavelengths for the positive and negative charges respectively. $\vec{r}_s$ and $\vec{r}_j$ are the position vectors for the positive ions and the negative ions respectively. In the partition function $Q$, the function $h(\vec{r})$ is introduced to generate the averaged charge density $<e\rho(\vec{r})>=\partial \ln Q/\partial h(\vec{r})|_{h=0}$. $\beta$ is the inverse temperature and the Hamiltonian $H$ is the sum of two terms by $H=H_C+H_N$. $H_C$ is the Coulomb energy from the EI, reading $H_C=\frac{e^2}{2}\int d\vec{r} d\vec{r}' \rho(\vec{r}) C(\vec{r},\vec{r}')\rho(\vec{r}')$.
By using the HST, we have
\begin{align}
\label{hc}
e^{-\beta H_c} =
\frac{1}{ Z_c}\int [\mathcal{D} \xi] e^{A_c}
\end{align}
with $A_c=- \frac{1}{2}\int dr\epsilon [\nabla_r \xi(r)][\nabla_{r}\xi(r)]-i\int dr \rho e \sqrt{\beta}\xi$ and $Z_c=\int [\mathcal{D} \xi] e^{- \frac{1}{2}\int dr dr' \xi(r')C^{-1} \xi(r)}$. Here, $\xi$ is an auxiliary field and $[\mathcal{D} \xi]$ is the measure of the field. To be self-contained, we present the derivation of Eq.(\ref{hc}) in Appendix (\ref{eqhc}). \\

The Hamiltonian for the NEIs reads $H_N=\frac{1}{2}\int d\vec{r} d\vec{r}' p(\vec{r}) D(\vec{r},\vec{r}')p(\vec{r}')$. To simplify our study, we have assumed that the function $D(\vec{r},\vec{r}')$ of the pairwise potential between any two ions is the same regardless of their species. We will specify the generalized potential function dependent on the species of ions in Subsection \ref{gener}. Now, we express $H_N$ in the reciprocal space by using the FT to get $H_N=\frac{1}{2} \int dG g^{\dagger} \lambda T g$. We discrete the reciprocal space with a reciprocal lattice by introducing an infinitesimal volume $\Delta$. In this way, we rewrite the integral of $H_N$ in the form of $H_N=\frac{1 }{2} \sum_G  (\sqrt{\Delta T}g)^{\dagger} \lambda (\sqrt{\Delta T}g)$. Note that $T$ is always positive according to our definition before. Here, $\sum_G$ means that the sum is over the lattice in the reciprocal space. Now we apply the HTS at each lattice site of $G$ to get
\begin{align}
\label{hn}
&e^{-\beta H_N}=e^{-\frac{1}{2} \sum_G  (\sqrt{\beta\Delta T}g)^{\dagger} \lambda (\sqrt{\beta\Delta T}g)}\nonumber\\
&=\frac{1}{Z_N}\int [\mathcal{D}\gamma^{\dagger}\mathcal{D}\gamma] e^{A_N}
\end{align}
with $A_N=-\frac{1}{2}\sum_G(\gamma^{\dagger}\gamma)+\frac{1}{2}\sum_G \sqrt{-\lambda} \sqrt{\beta\Delta T}(\gamma g+\gamma^{\dagger}g^{\dagger})$ and $Z_N=\int [\mathcal{D}\gamma^{\dagger}\mathcal{D}\gamma] e^{-\frac{1}{2}\sum_G(\gamma^{\dagger}\gamma)}$. Here, $\gamma$ and $\gamma^{\dagger}$ are the auxiliary fields. $[\mathcal{D}\gamma^{\dagger}\mathcal{D}\gamma]$ are the measures of the auxiliary fields. Note that $\sqrt{-\lambda}$ could be $+1$ or $i$ since $\lambda$ takes the values of $-1$ or $+1$, and is functional of $G$.\\

We substitute Eq.(\ref{hc}) and Eq.(\ref{hn}) into Eq.(\ref{Q}) to rewrite the partition function $Q$. For convenience, we use $l_B=4\pi l_B'$ to scale all the lengths in this study. $l_B'$ is the Bjerrum length defined by $l_B'=e^2\beta/(4\pi\epsilon)$, which is $6.96\AA$ for water at the room temperature. We rewrite $\vec{r}/l_B$ by $r$, $e\sqrt{\beta}\xi$ by $\xi$, $eh$ by $h$, $Gl_B$ by $G$, $\Delta l_B^3$ by $\Delta$, $\beta T/l_B^3$ by $T$, $\lambda_{\pm}/l_B$ by $\lambda_{\pm}$, and $\sigma l_B^3$ by $\sigma$ to save symbols. Then, the partition function $Q$ expressed by the fields reads
\begin{align}
Q&=\frac{1}{Z_cZ_N}\frac{1}{N_+!N_-!\lambda_+^{3N_+}\lambda_-^{3N_-}}\int [\mathcal{D}\gamma^{\dagger}\mathcal{D}\gamma ][\mathcal{D} \xi]\times\nonumber\\
&e^{A_1+A_2+A_3}\times \Lambda_+^{N_+}\Lambda_-^{N_-}
\end{align} 
with
\begin{align}
&A_1=- \frac{1}{2}\int dr~[\nabla_r \xi(r)][\nabla_{r}\xi(r)],\nonumber\\
&A_2=-\frac{1}{2}\sum_G(\gamma^{\dagger}\gamma),\nonumber\\
&A_3=\int d r~(-i\xi+h)\sigma,\nonumber\\
&\Lambda_{\pm}=\int dr~ e^{B_{\pm}+\Gamma},\nonumber\\
&B_{\pm}=(-i\xi+h)z_{\pm},\nonumber\\
&\Gamma=\frac{1}{2}\sum_G\omega\sqrt{\Delta} [\gamma e^{-iGr}+\gamma^{\dagger}e^{iGr}].\nonumber
\end{align}
For clarity, we use the notation $\omega=\sqrt{-\lambda T}$. We introduce the fugacity $\mu_{+}$ for positive charges and $\mu_{-}$ for negative ones. The grand canonical partition function is $\Xi=\sum_{N_+=0}^{\infty}\sum_{N_-=0}^{\infty}Q(N_+,N_-)\mu_+^{N_+}\mu_-^{N_-}$, which reads
\begin{align}
\label{Xi}
\Xi=&\frac{1}{Z_cZ_N}\int [\mathcal{D}\gamma^{\dagger}\mathcal{D}\gamma ][\mathcal{D} \xi] e^{A_1+A_2+A_3+A_4+A_5}
\end{align}
with $A_4=w_+\Lambda_+$, $A_5=w_-\Lambda_-$, and $w_{\pm}=\mu_{\pm}/\lambda_{\pm}^3$. For convenience, $\Xi$ can be written as $\Xi=\Xi_{N}\cdot \Xi_{c}$ with
\begin{align}
&\Xi_{N}=\frac{1}{Z_N}\int [\mathcal{D}\gamma^{\dagger}\mathcal{D}\gamma]e^{A_2},\nonumber\\ &\Xi_{c}=\frac{1}{Z_c}\int [\mathcal{D} \xi]e^{A_1+A_3+A_4+A_5}.\nonumber
\end{align}
We will show later that $\Xi_N$ defines Gaussian random fields for the NEIs.\\

\subsection{saddle point solution}
By functional derivative, we get the saddle point solution of the auxiliary filed $\xi$, which satisfies the equation $\partial \Xi/\partial \xi|_{h=0}=0$. Explicitly, it is 
\begin{align}
\label{NEIPB}
-\nabla_r^2[i\xi]=\sigma+z_+w_+e^{-z_+[i\xi]+\Gamma}+z_-w_-e^{-z_-[i\xi]+\Gamma}.
\end{align}
We apply $<\rho(r)>=\partial \Xi/\partial h|_{h=0}$ to get 
\begin{align}
\label{cdens}
<\rho(r)>=<\sigma+z_+w_+e^{-z_+[i\xi]+\Gamma}+z_-w_-e^{-z_-[i\xi]+\Gamma}>
\end{align}
for the charge density, which is the averaged value of the right hand side in Eq.(\ref{NEIPB}).\\

If we drop off the term $\Gamma$ which is originated from the NEIs, Eq.(\ref{NEIPB}) recovers the mean field PB equation since the right hand side of Eq.(\ref{NEIPB}) is exactly the charge density according to Eq.(\ref{cdens}). We understand that $[i\xi]$ is the electrostatic field in the charged liquid. Generally, the auxiliary field $\xi$ is complex due to the ion fluctuations by the EI. We write $\xi=\xi_R-i\xi_I$ with $-i\xi_I$ the mean field of $\xi$ and $\xi_R$ the fluctuations around $-i\xi_I$. Thus, $\xi_I$ represents the electrostatic field according to our understanding of $[i\xi]$ and $\xi_R$ is originated from the ion fluctuations. To study the ion-ion correlations in the charged liquids, the contribution from $\xi_R$ has to be considered to go beyond the mean field PB equation. By following the work in Ref.\cite{Wan1}, we will transform the ion fluctuations to a random field and implement the random field in a modified PB equation, which will be performed in Subsection \ref{EIRF}.\\

The coupling between $\xi_I$ and $\xi_R$ will be much more complicated if $\Gamma$ is involved. Generally, $\Gamma$ is complex, meaning that the NEIs influence not only the ion fluctuations $\xi_R$ but also the electrostatic field $\xi_I$. It is very difficult to calculate $\Gamma$ directly. In the following, we will express the term $\Gamma$ by random fields. In this way, it is much easier to generate random numbers for $\Gamma$ than calculate $\Gamma$ directly.\\

\subsection{random field $\Gamma$ for NEIs}
We denote $\theta=\gamma/\sqrt{\Delta}$, and write $\Gamma$ in the integral form by using $\sum_G \Delta =\int dG$. In this way, we have $\Gamma=\frac{1}{2}\int dG~\omega[\theta e^{-iGr}+\theta^{\dagger}e^{iGr}]$. By replacing $\gamma=\theta \sqrt{\Delta}$ in $A_2$, we have $A_2=-\frac{1}{2}\int dG~ \theta^{\dagger}\theta$. For clarity, we write $\theta=a+ib$ with $a$ and $b$ being real. Then, we obtain $\Gamma=\int dG~\omega[a\cos(Gr)+b\sin(Gr)]$ and $A_2=-\frac{1}{2}\int dG~(a^2+b^2)$. The measure $[\mathcal{D}\gamma^{\dagger}\mathcal{D}\gamma ]$ in $\Xi_N$ then is transformed to the measure $[\mathcal{D}a\mathcal{D}b]$. Therefore, the measure $[\mathcal{D}a\mathcal{D}b]$ and the term $e^{A_2}$ in $\Xi_N$ interpret that $a$ and $b$ are Gaussian random fields, which can be understood as the spatial noises mapped from the temporal noise in the theory of stochastic process. The expression $e^{A_2}$ reveals that the Gaussian random fields $a$ and $b$ both have the expectation equaling zero and the variance equaling $1$. The Gaussian distribution of the random fields $a$ and $b$ is denoted by $N(0,1)$. Since $a$ and $b$ are Gaussian random fields, $a\cos(Gr)+b\sin(Gr)$ is also a Gaussian random field because of $\cos^2(Gr)+\sin^2(Gr)=1$. We introduce a Gaussian random variable $\vartheta=a\cos(Gr)+b\sin(Gr)$ to simplify the notation, and have
\begin{align}
\label{B1}
\Gamma=\int dG~\omega \cdot \vartheta=\int dG~\sqrt{-\lambda T} \cdot \vartheta,
\end{align}
with $\vartheta$ following the Gaussian distribution $N(0,1)$. Note that the definition of $\vartheta$ indicates that $\vartheta$ is functional of $G$ and $r$. After the integration over the reciprocal space $G$, $\Gamma$ is the random field defined in the real space of $r$.\\

For convenience, we call the random field of the NEIs by NEIRF, which has been denoted by $\Gamma$. Generally, $\Gamma$ is complex, and can be expressed as $\Gamma=\Gamma_R+i\Gamma_I$ by two real values of $\Gamma_R$ and $\Gamma_I$. Suppose the real part of $\sqrt{-\lambda T}$ in Eq.(\ref{B1}) is denoted by $V_R'$ and the imaginary part by $V_I'$. Both $V_R'$ and $V_I'$ are functional of $G$. We further denote $V_R^2=\int dG~(V_R')^2$ and $V_I^2=\int dG~(V_I')^2$. Since $\vartheta$ is a Gaussian random field, $\Gamma_R$ and $\Gamma_I$ both are Gaussian random fields according to the probability theory. $\Gamma_R$ follows the Gaussian distribution $N(0,V_R^2)$ with the averaged value of $0$ and the variance of $V_R^2$. $\Gamma_I$ follows the Gaussian distribution $N(0,V_I^2)$. In this way, we can generate the random fields $\Gamma$ by the Gaussian distributions directly in the calculations. The random fields of  $\Gamma_R$ and $\Gamma_I$ are assumed to be independent. The joint probability density of  $\Gamma_R$ and $\Gamma_I$ reads 
\begin{align}
\label{prodens}
f=\frac{1}{2\pi V_RV_I}e^{-\frac{1}{2}(\Gamma_R^2/V_R^2+\Gamma_I^2/V_I^2)}
\end{align}
which will be checked in the subsection \ref{noisepattern}. \\

We substitute Eq.(\ref{B1}) into Eq.(\ref{NEIPB}) to get a modified PB equation with the NEIRF. The solution to the equation reflects the effects of the NEIs in the charged liquid. However, the equation still has one ingredient missing, which is the ion fluctuations by the EI for the ion correlation. 

\subsection{random field $\eta$ for EI}
\label{EIRF}
We write $\xi$ by two components $\xi=\phi-i\Psi$. Here, $-i\Psi$ is close to the mean field solution to Eq.(\ref{NEIPB}) with $\Gamma$ involved, but is not exactly equivalent to the mean field solution due to the modification by the ion fluctuations in the charged liquid. And $\phi$ fluctuates around $-i\Psi$, and is responsible for the ion fluctuations. The measure $[\mathcal{D}\xi]$ in $\Xi_c$ then is transformed to the measure $[\mathcal{D}\phi]$ by shifting $-i\Psi$. We assume that the ion fluctuation is not intensive leading to a small field $\phi$. This assumption is reasonable because the charged liquid we study is at the equilibrium and not in flowing. In this way, the component $\Lambda_{\pm}$ in $\Xi_c$ can be expanded to the second order by $\phi$, which reads
\begin{align}
\Lambda_{\pm}\approxeq \int dr~e^{-z_{\pm}\Psi +\Gamma}(1-iz_{\pm}\phi -\frac{1}{2}z_{\pm}^2\phi^2).
\end{align}
For the expansion, we have set $h=0$. The exponent $A_1+A_3+A_4+A_5$ in $\Xi_c$ is rewritten as a sum of three terms, and $\Xi_c$ is rewritten as
\begin{align}
\label{xic}
\Xi_c=\frac{1}{Z_c}\int [\mathcal{D}\phi]e^{\int dr[B_1+i\phi B_2-\frac{1}{2}\phi^2 B_3]}
\end{align}
with
\begin{align}
&B_1=\frac{1}{2}[(\nabla_r \Psi)^2-(\nabla_r \phi)^2]-\Psi \sigma+w_+e^{-z_+\Psi+\Gamma}+w_-e^{-z_-\Psi+\Gamma},\nonumber\\
&B_2=-\nabla_r^2 \Psi-\sigma-z_+w_+e^{-z_+\Psi+\Gamma}-z_-w_-e^{-z_-\Psi+\Gamma},\nonumber\\
&B_3=z_+^2w_+e^{-z_+\Psi+\Gamma}+z_-^2w_-e^{-z_-\Psi+\Gamma}.\nonumber
\end{align}
The first term $\nabla_r^2 \Psi$ in $B_2$ is obtained from $\nabla_r \phi \nabla_r \Psi$ in the term $A_1$ by using the integration by parts. The operator $\nabla_r$ has been specified before and is applied in the real space of $r$. Without confusion, we drop off the subscript $r$ and write $\nabla_r$ by $\nabla$ for simplicity. 
\\

We discrete the real space $r$ with a lattice by introducing an infinitesimal volume $\tau$ and express the integral in the form of sum by $\int dr=\sum_r \tau$. Here, $\sum_r$ means that the sum is over the lattice in the real space $r$. We apply the discretization on the component of $B_3$ in $\Xi_c$ of Eq.(\ref{xic}) and have
\begin{align}
\label{EIHST}
&e^{-\frac{1}{2}\int dr\phi^2 B_3}=e^{-\frac{1}{2}\sum_r (\sqrt{B_3 \tau}\phi)(\sqrt{B_3\tau}\phi) }\nonumber\\
&=\frac{1}{Z_{\alpha}}\int [\mathcal{D}\alpha]e^{-\frac{1}{2}\sum_r\alpha^2}e^{i\sum_r\alpha \sqrt{B_3\tau}\phi}
\end{align} 
with $Z_{\alpha}=\int [\mathcal{D}\alpha]e^{-\frac{1}{2}\sum_r\alpha^2}$. For the second equality in the above equation, the HST has been applied with $\alpha$ introduced. $\alpha$ is $r$ dependent and defined at each lattice site in the real space. We define $\eta=\alpha/\sqrt{\tau}$ and use the transform $\sum_r \tau=\int dr$ to re-express the second equality in Eq.(\ref{EIHST}), leading to
\begin{align}
\label{eta}
e^{-\frac{1}{2}\int dr\phi^2 B_3}=\frac{1}{Z_{\eta}}\int [\mathcal{D}\eta]e^{-\frac{1}{2}\int dr\eta^2}e^{i\int dr\eta \sqrt{B_3}\phi}
\end{align}
with $Z_{\eta}=\int [\mathcal{D}\eta]e^{-\frac{1}{2}\int dr\eta^2}$. In the above equation, we have the factor $e^{-\frac{1}{2}\int dr\eta^2}$ and the measure $[\mathcal{D}\eta]$, which means that $\eta$ can be interpreted as a Gaussian random field and follows the Gaussian distribution $N(0,1)$. \\

Substituting Eq.(\ref{eta}) into Eq.(\ref{xic}), we have
\begin{align}
\label{xic1}
\Xi_c=\frac{1}{Z_cZ_{\eta}}\int [\mathcal{D}\phi][\mathcal{D}\eta]e^{-\frac{1}{2}\int dr\eta^2}e^{\int dr[B_1+i\phi (B_2+\eta \sqrt{B_3})]}.
\end{align}
Considering that the fluctuation $\phi$ is small, we neglect the term $(\nabla \phi)^2$ in the factor $B_1$. The functional integral of $[\mathcal{D}\phi]$ in Eq.(\ref{xic1}) leads to a functional $\delta(B_2+\eta \sqrt{B_3})$ in the real space of $r$.
Therefore, we obtain an equation of $B_2+\eta \sqrt{B_3}=0$. This equation can be generalized to a charged liquid containing various ionic species $(k=1,2,\dots)$, and is written explicitly as
\begin{align}
\label{finalequ}
-\nabla^2 \Psi=\sigma+e^{\Gamma}\sum_kz_kw_ke^{-z_k\Psi}+\sqrt{e^{\Gamma}\sum_kz_k^2w_ke^{-z_k\Psi}}\cdot \eta
\end{align}
with $z_k$ the charge value of the $k$-th ionic species. Eq.(\ref{finalequ}) is the main result in this study. The full solution to Eq.(\ref{finalequ}) contains both the informations of the NEIs and the ion fluctuations of the EI. If we set $\Gamma=0$ and $\eta=0$, Eq.(\ref{finalequ}) is reduced to be the mean field PB equation. If we keep $\eta$ but set $\Gamma=0$, then the NEIs are removed and only the ion fluctuations are involved in the equation.\\

The grand canonical partition function $\Xi$ in Eq.(\ref{Xi}) contains the statistical physics in the charged liquid. We need figure out the physical meanings of the quantities obtained from $\Xi$. Generally, $\Gamma$ is a complex field, which is originated from the term $\sqrt{-\lambda T}$ in Eq.(\ref{B1}). The FT components with $\lambda=+1$ obtained from the FT on the NEIs $D(\vec{x})$ contribute to the imaginary part of $\Gamma$, while the components with $\lambda=-1$ contribute to the real part of $\Gamma$. We are not able to identify the interactions of the FT components with $\lambda=+1$ as the attractive or repulsive interactions, but recognize that the interactions of $\lambda=+1$ are opposite to the interactions if $\lambda=-1$. Therefore, the complex nature of $\Gamma$ is due to the opposite interaction of the NEIs. We use the phrase of interaction phase for $\lambda$. In the charged liquid, the attractive or the repulsive interactions between ions occur randomly. It is expected that the interaction phase of the NEIs should be averaged to be zero. In the numerical calculations, the averaged value of the interaction phase will be very small, which will be checked later.\\

Eq.(\ref{finalequ}) is a complex differential equation due to the complex field $\Gamma$. The solution $\Psi$ then is a complex field. The electrostatic field is real, and can be obtained by $(\Psi+\Psi^{\dagger})/2$. Thus, the real part of $\Psi$ is assigned to the electrostatic field and the imaginary part of $\Psi$ is responsible for the interaction phase by the NEIs.\\ 

\subsection{charge conservation}
Similarly, $<\rho(r)>$ in Eq.(\ref{cdens}) is a complex function. The real part of $w_ke^{-z_k\Psi+\Gamma}$ in Eq.(\ref{cdens}) is the number density of the $k$-th ionic species and the imaginary part is responsible for the interaction phase by the NEIs. In the charged liquid, the distribution of ions should not change the total ion number of each ionic species. And the charge of each ionic species should be conserved~\cite{Wan2,Liu}. Such charge conservation must be implemented in Eq.(\ref{finalequ}). We denote the real part of a complex value $\mathbb{C}$ by $Re[\mathbb{C}]$, and denote the averaged number density of the $k$-th ionic species in the computational domain by $M_k$. The charge conservation requires that $Re[\int dr w_ke^{-z_k\Psi+\Gamma}]=\int dr M_k$, leading to
\begin{align}
\label{wk}
w_k=\frac{M_k\int dr}{Re[\int dr~ e^{-z_k\Psi+\Gamma}]}.
\end{align}
The integrations are over the total volume of the charged liquid.\\ 

\subsection{boundary conditions}
Before we solve Eq.(\ref{finalequ}), a boundary condition(BC) for the EI should be applied. There are three basic BCs for the equation~\cite{Wan2}. The first BC is the Dirichlet BC, in which the electrostatic potentials are fixed at the boundaries. For example, we apply voltage drops on solid boundaries of the charged liquid. The voltages at the solid boundaries are the Dirichlet BCs. The second BC is the Neumann BC, in which external charges are fixed in the solid boundaries. In the Neumann BC, the derivatives of the electrostatic potentials with respect to the spatial coordinates should equal the densities of the fixed external charges through the Gauss's law. The last BC is the Robin BC, which is the mixture of the Dirichlet BC and the Neumann BC. Besides the three basic BCs mentioned above, there exist several other BCs, such as the charge regulation model and the potential trap model~\cite{Wan2,5Camara,5Carnie,5Behrens,5Gentil}. In this study, we use the Dirichlet BC to demonstrate our theory.\\

We note that such BC is for the EI instead of the NEIs. That means the ions in the charged liquid do not feel the NEIs from the solid boundaries. In order to catch the physics of the ions close to the boundaries, such as the stern layer due to the steric effect, we need implement additional BCs for the NEIs. This is a very interesting question, but beyond the present study. In this study, we apply the Dirichlet BC for the EI, and only show the NEIs in the bulk domain for the charged liquid.\\

\subsection{algorithm}
\label{algorithm}
To solve a real differential equation with random fields, Path Integral method combined with Monte Carlo technique (PIMC) has been a well defined method for the solving~\cite{Wan1}. However, the PIMC fails in this study since Eq.(\ref{finalequ}) is a complex differential equation. After testing, we find that it is possible to solve Eq.(\ref{finalequ}) by using the finite element method (FEM). We also find that the FEM is feasible for the charged liquid with the total net charge neutral, and the FEM is not converged if the total net charge is nonzero. Therefore, a general mathematical tool to solve Eq.(\ref{finalequ}) is still lacking. In this study, we will use the FEM to study the charged liquid consisting of two ionic species with opposite charges and the total net charge of the liquid is neutral.\\

In the following, we list the algorithm for the solving of Eq.(\ref{finalequ}). 
\begin{itemize}
\item{Step 1:\\
Make the FT on the pairwise function $D(\vec{x})$ of the NEIs to get $\lambda$ and $T$ which are functional of $G$. Calculate the real part $V_R'$ and the imaginary part $V_I'$ of $\sqrt{-\lambda T}$ respectively. Get $V_R^2=\int dG~(V_R')^2$ and $V_I^2=\int dG~(V_I')^2$.}
\item{Step 2:\\
Generate the random numbers $\Gamma_R$ and $\Gamma_I$ according to the probability density Eq.(\ref{prodens}) with $V_R$ and $V_I$ obtained in step 1. Get $\Gamma=\Gamma_R+i\Gamma_I$ at each lattice site in the real space of $r$. Generate the random numbers $\eta$ according to the Gaussian probability density $N(0,1)$ at each lattice site of the real space $r$. Now Eq.(\ref{finalequ}) is ready. The term of the random number $\eta$ is treated in stratonovich sense.}
\item{Step 3:\\
Apply the FEM to solve Eq.(\ref{finalequ}) to get one solution $\Psi$.  }
\item{Step 4:\\
Repeat the calculations from step 2 to step 3 to get many solutions of $\Psi$. Finally make the statistical average over the solutions of $\Psi$ for the final results.}
\end{itemize}
To solve Eq.(\ref{finalequ}) with the FEM, the iteration method is applied. In details, $\Psi$ is initialized to be $\Psi_0$. The right hand side of Eq.(\ref{finalequ}) then is obtained as $RHS(\Psi_0)$ with $\Psi_0$ input. By solving the equation $-\nabla^2\Psi_1=RHS(\Psi_0)$ with the FEM, the solution $\Psi_1$ is obtained as the input to $RHS(\Psi_1)$ for the next iteration. For the FEM, the software FreeFem++ is applied~\cite{Freefem}. \\

\subsection{generalization}
\label{gener}
In this study, we focus on the charged liquid in which the function $D(\vec{x})$ of the NEIs between any two ions is the same for simplicity, even if between two different species. The NEIRF noted by $\Gamma$ in Eq.(\ref{finalequ}) is shared by all the ions regardless of their species. Actually, we can generalize the NEIRF to the charged liquids with the NEIRF dependent on the ionic species. The generalized version of Eq.(\ref{finalequ}) has been indicated in Eq.(\ref{generalizedequ}) in Appendix \ref{generalization} for the readers' interests. We emphasize that the present study is still focused on Eq.(\ref{finalequ}).\\

\section{results}
The charged liquid we study is one dimensional, starting from $r=0$ to $r=L$. For convenience, we denote the boundary at $r=L$ by BD1, and the boundary at $r=0$ by BD2. The liquid contains two ionic species with opposite charges. We set $z_+=+1$ and $z_-=-1$ in this study and the number density $M=M_+=M_-$ for the charge neutrality in the liquid. The Dirichlet BC is applied by setting $\Psi(r=L)=2$ at the BD1 and $\Psi(r=0)=0$ at the BD2.\\ 

The NEIs are general in this study. Here, we focus on the steric effect of the ions in the liquid to demonstrate our theory. The steric effect is realized by taking the hard core model for the ions. For simplicity, all the ions have the same radius $R$ and the repulsive potential energy $h$. Explicitly, the potential function $D(x)$ is zero if the distance $x$ between any two ions is larger than $2R$. And $D(x)=h$ if $x<2R$. In the numerical calculations, it is impossible to take $h=\infty$ for the hard core model. We will vary the value of $h$ to show the steric effect.\\

\subsection{random field $\Gamma$}
\label{noisepattern}
By applying the FT on the function $D(x)$, we have an analytic expression of $\lambda T=h\sin(GR)/(8\pi^4G)$. We fix $R=0.05$ which is close to the size of one water molecular. \\

$\Gamma_R$ and $\Gamma_I$ are random numbers, which can be calculated through Eq.(\ref{B1}). In details, the random numbers $\vartheta$ are generated at each lattice site in the reciprocal space of $G$ according to the Gaussian distribution $N(0,1)$ for one trial. And then make the integration by Eq.(\ref{B1}) in Stratonovich sense to get one pair of $\Gamma_R$ and $\Gamma_I$, since the function $\lambda T$ is ready. After many trials, we obtained many pairs of the random numbers $\Gamma_R$ and $\Gamma_I$, and then figure out the probability density of the pairs of the random numbers. We present the numerical result in Fig.1(a) with $h=100$ for the illustration. The color bar in Fig.1(a) is for the joint probability density(JPD) of $\Gamma_R$ and $\Gamma_I$. We find that the JPD can be approximated to be Gaussian in Fig.1(a).\\

We assume that $\Gamma_R$ and $\Gamma_I$ are independent to each other, and the analytical JPD then can be derived from Eq.(\ref{B1}) directly by using the probability theory. According step 1 in subsection \ref{algorithm}, we calculate $V_R=\sqrt{h}\times 0.479$ and $V_I=\sqrt{h}\times 0.623$. The analytical JPD shows that $\Gamma_R$ and $\Gamma_I$ follow the Gaussian distributions $N(0,V_R^2)$ and $N(0,V_I^2)$ respectively, which has been indicated in Eq.(\ref{prodens}). The assumption of the independence between $\Gamma_R$ and $\Gamma_I$ benefits the numerical calculations. We can generate $\Gamma_R$ and $\Gamma_I$ directly from Eq.(\ref{prodens}) instead of Eq.(\ref{B1}). In the following, we verify the assumption with $h=100$.\\

In Fig.1(b), the scattered data are from Fig.1(a) along the path of $\Gamma_I=0$. The solid line in Fig.1(b) is obtained by Eq.(\ref{prodens}) with $\Gamma_I=0$ and $V_R=4.79$. It indicates that the solid line fits the scattered data in Fig.1(b) very well. In Fig.1(c), the scattered data are from the path of $\Gamma_R=0$ of Fig.1(a), which can be fit by Eq.(\ref{prodens}) with $\Gamma_R=0$ and $V_I=6.23$. In Fig.1(d), we plot the scattered data chosen from the path of $\Gamma_I=\Gamma_R$ in Fig.1(a), and plot the solid line by $JPD=e^{-(\sqrt{2}\Gamma_I)^2/(2V_{RI}^2)}/(2\pi V_RV_I)$ with $V_{RI}^2=2V_R^2V_I^2/(V_R^2+V_I^2)$ according to Eq.(\ref{prodens}). The matching between the solid line and the scattered data can be found in Fig.1(d). The results in Fig.1 confirm that the random fields $\Gamma_R$ and $\Gamma_I$ can be treated to be independent to each other and can be generated directly by Eq.(\ref{prodens}) for the following numerical calculations.\\

\begin{figure}[tbh!]
\includegraphics[scale=0.6]{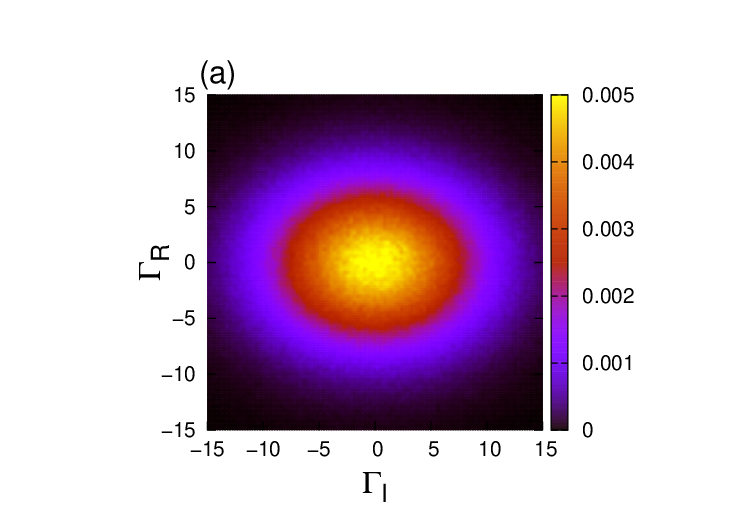}
\includegraphics[scale=0.2]{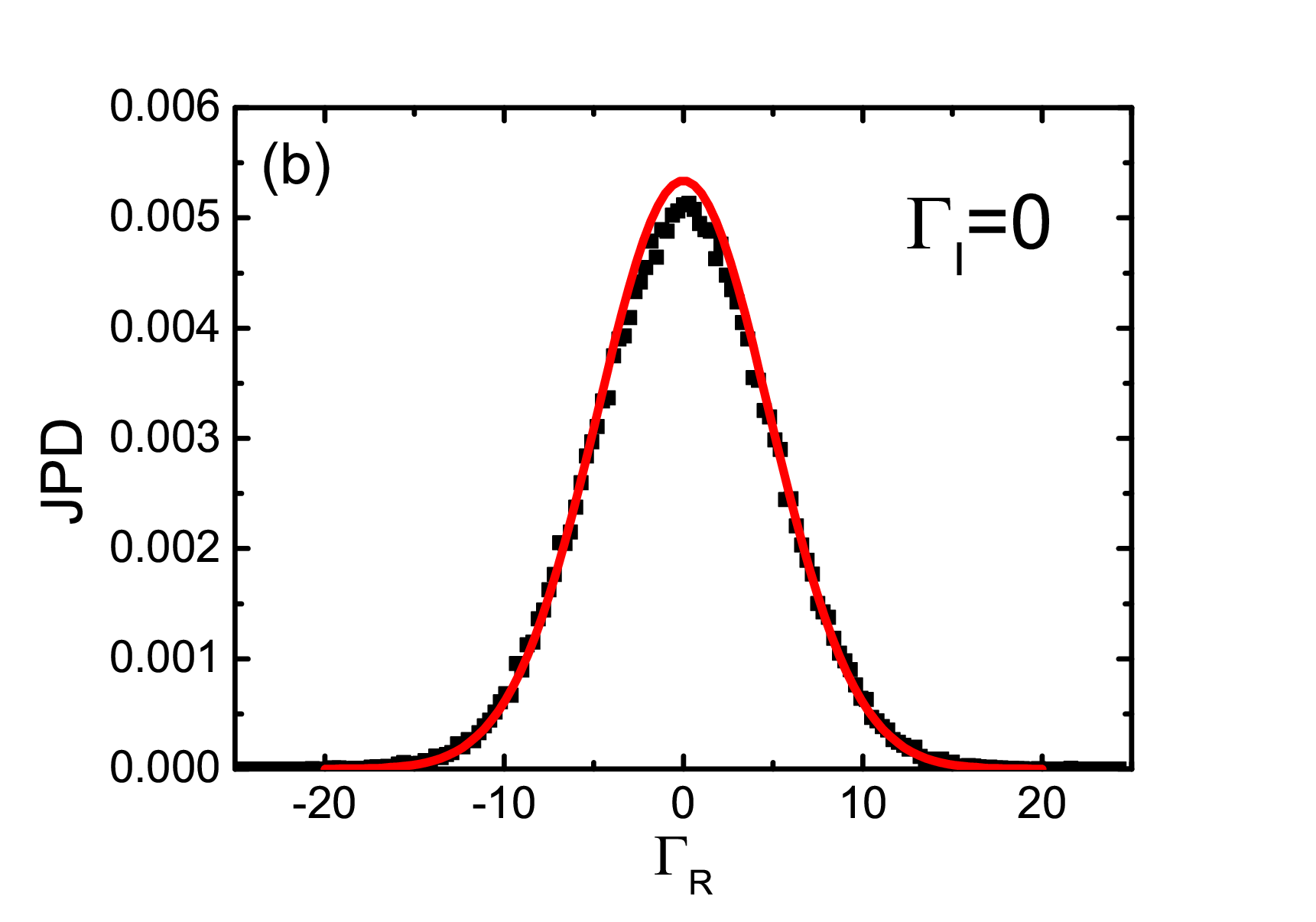}
\includegraphics[scale=0.2]{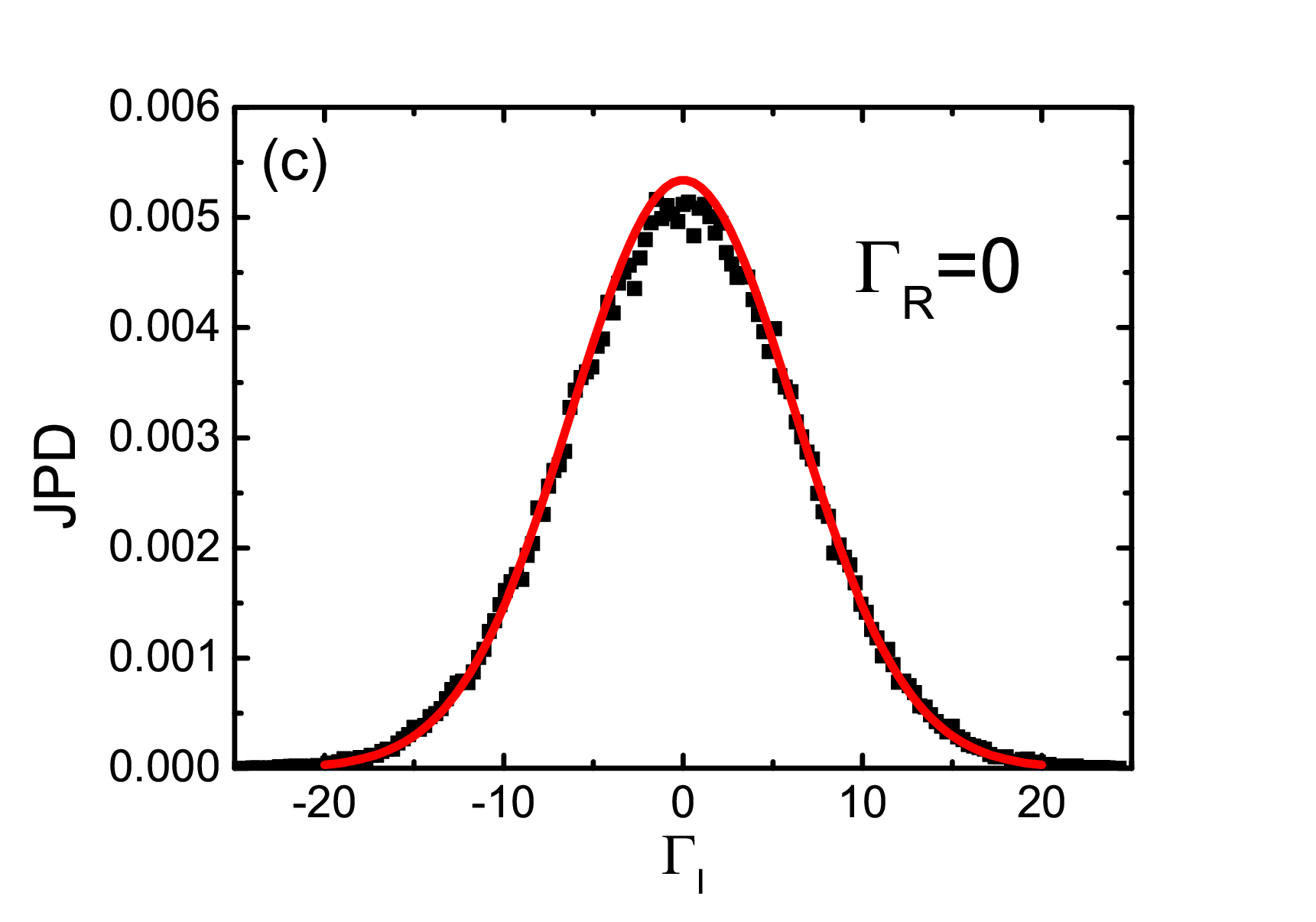}
\includegraphics[scale=0.2]{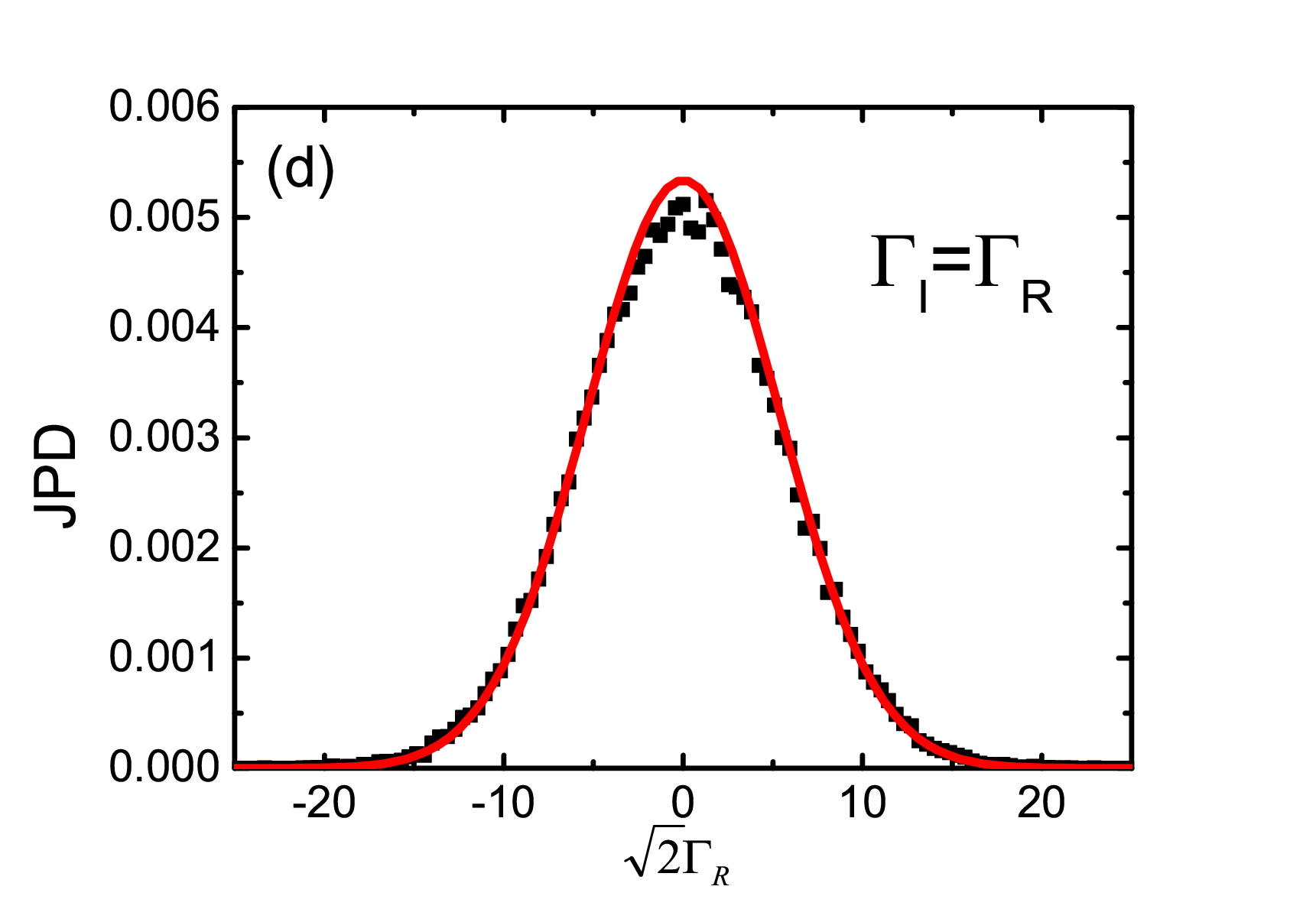}
\caption{The joint probability density(JPD) for the random fields $\Gamma_R$ and $\Gamma_I$.(a) The JPD of $\Gamma_R$ and $\Gamma_I$ is calculated by Eq.(\ref{B1}) directly. (b) The scattered data are obtained from the path of $\Gamma_I=0$ in (a). (c)The scattered data are obtained from the path of $\Gamma_R=0$ in (a). (d) The scattered data are obtained from the path of $\Gamma_I=\Gamma_R$ in (a). All the scattered data in (b),(c) and (d) can be fit by the solid lines obtained from Eq.(\ref{prodens}).}
\end{figure}

\subsection{steric effect}
After generating the random fields $\Gamma_R$, $\Gamma_I$ and $\eta$ according to their own distributions of the probability densities, Eq.(\ref{finalequ}) becomes a regular complex differential equation, and can be solved by the FEM to get $\Psi$. Considering the random nature of $\Psi$, we need generate several trials of $\Gamma$ and $\eta$ to get various $\Psi$, and make the statistical average over $\Psi$ for the final result. In order to study the fluctuation of the electrostatic field, we define the quantity $\Psi_{var}=\sqrt{E(\Psi_{real}^2)-E(\Psi_{real})\times E(\Psi_{real})}$. Here, $E(\cdots)$ means the expectation of $\cdots$. $\Psi_{real}$ is referred to the real part of $\Psi$ which is the electrostatic field. In this study, we set the charge density $M=0.04$, by which the Debye length is calculated to be $5$. The solutions to Eq.(\ref{finalequ})are presented in Fig.2 with the length of the liquid set to be $L=20$, which is much larger than the Debye length. In this case, the Debye screening domains of the both boundaries do not intersect.\\

The real part of $\Psi$ is shown in Fig.2(a). In order to distinguish the results, we have enlarged one scale of the results in the insert figure. The line indicated by PB is the solution to the mean field PB equation, which is reduced from Eq.(\ref{finalequ}) with $\Gamma=0$ and $\eta=0$. The line indicated by ERPB is the solution to Eq.(\ref{finalequ}) by setting $\Gamma=0$ and still keeping $\eta$ as the random field of the EI. The ERPB result considers only the ion fluctuation of the EI and remove the NEIs from the charged liquid. It shows that the PB line and the ERPB line almost overlap in Fig.2(a) and can be distinguished in the insert figure, meaning that the ion-ion correlations due to the ion fluctuations really can modify the PB result.\\

The full solution to Eq.(\ref{finalequ}) with $\Gamma$ and $\eta$ involved contains the effects of the NEIs, which is denoted by NEPB for convenience. In order to show the steric effect, we have varied $h$ for the NEPB and indicated the results of the NEPB by various $h$ in the insert figure in Fig.2(a).  
\begin{figure}[tbh!]
\includegraphics[scale=0.25]{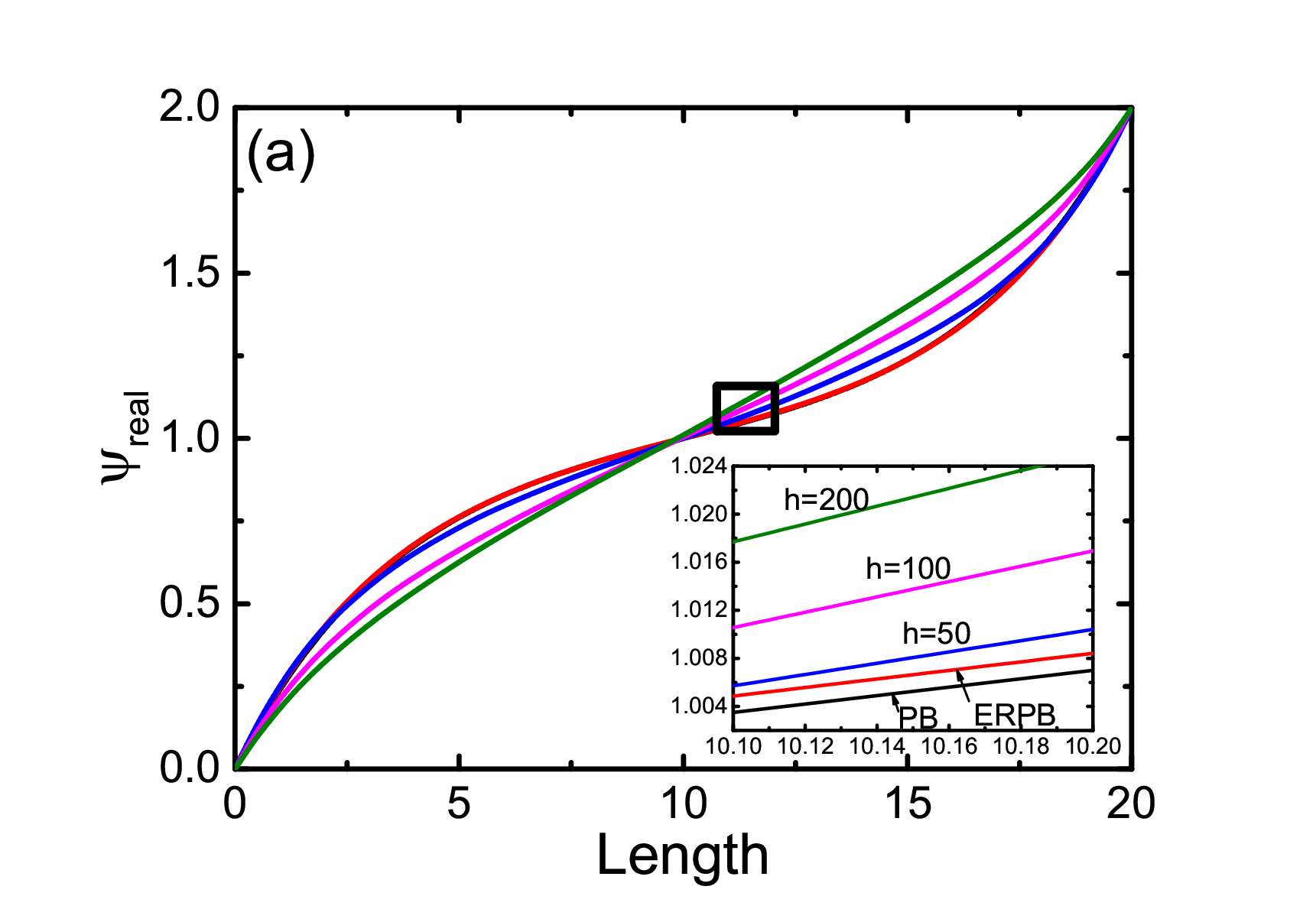}
\includegraphics[scale=0.25]{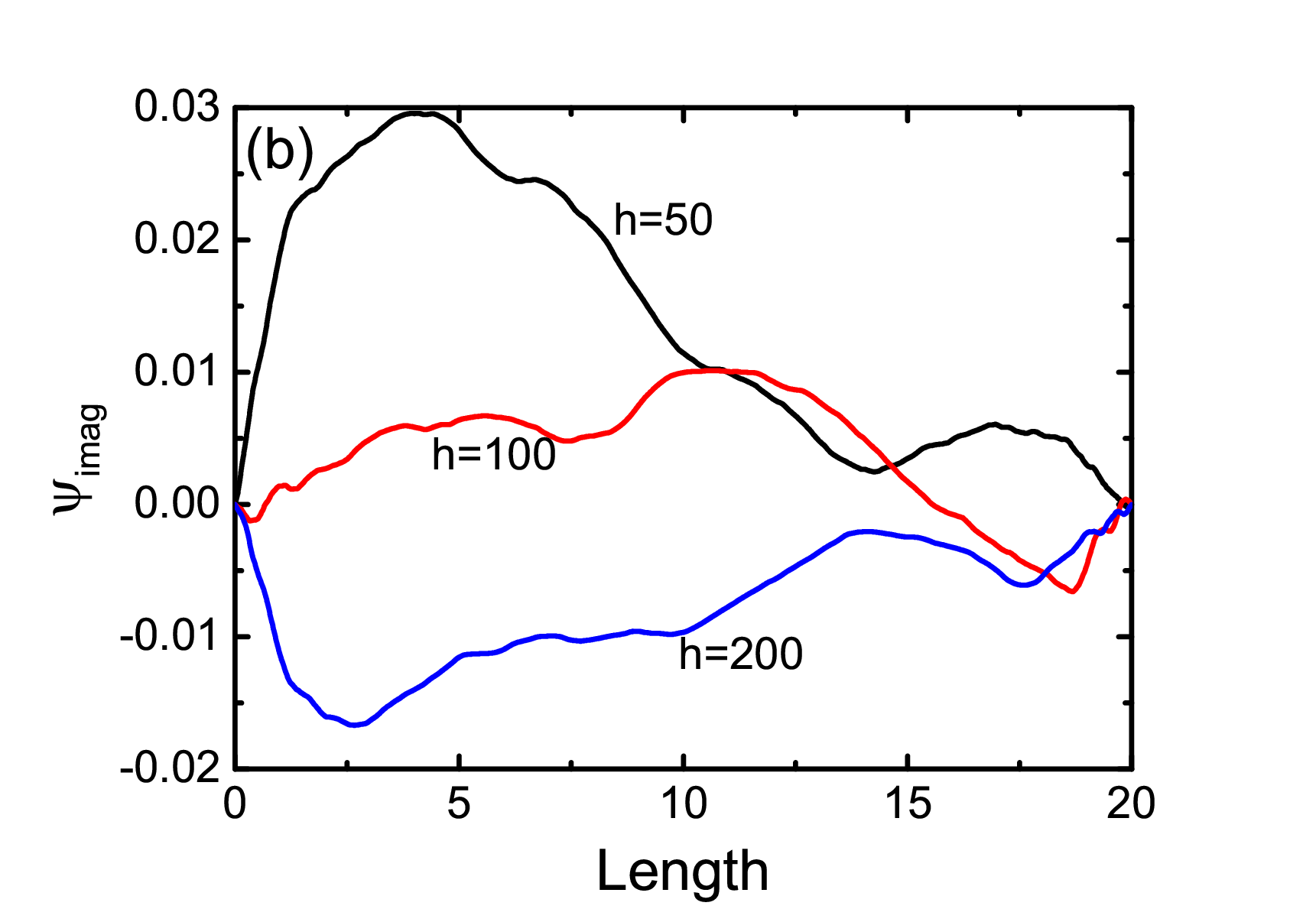}
\includegraphics[scale=0.25]{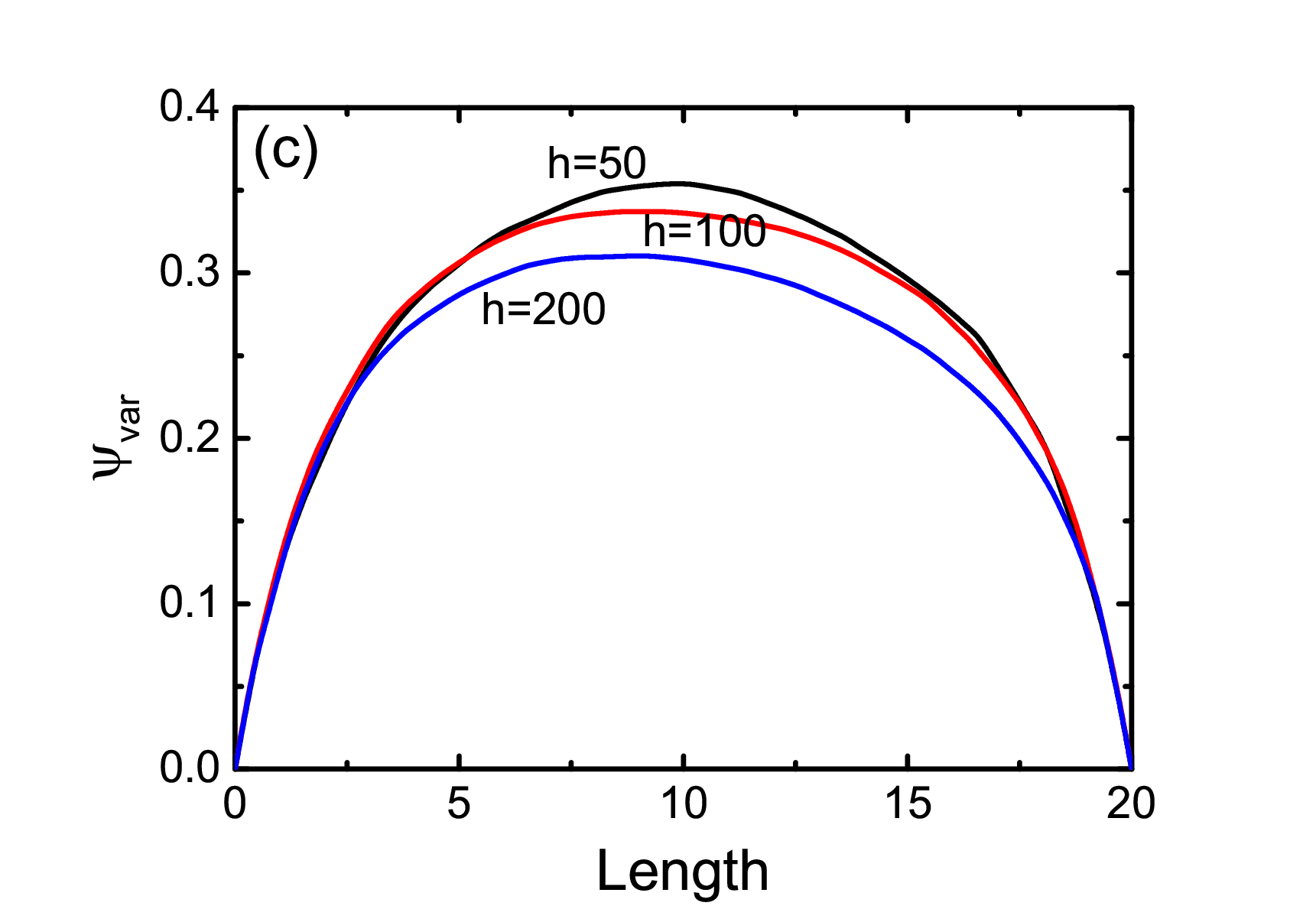}
\caption{Solutions of $\Psi$ to Eq.(\ref{finalequ}).(a)The real part of $\Psi$ is modified by the steric effect. (b) The imaginary part of $\Psi$ is close to zero. (c) The variance of $\Psi$ decreases with the increase of $h$.}
\end{figure}
It indicates that the NEPB lines deviate from both the PB line and the ERPB line, meaning that the NEIs play an important role in modifying the electrostatic field. With a larger value of $h$, the ions are much harder and the steric effect is stronger, which leads to a larger deviation. It is observed that for $h=200$, the electrostatic potential $\Psi_{real}$ is almost linear, showing that the Debye screening domain has been modified by the steric effect.\\

The imaginary part of $\Psi$ that is denoted by $\Psi_{imag}$ for various $h$ has been shown in Fig.2(b). It seems that $\Psi_{imag}$ is very small and close to zero when compared to $\Psi_{real}$. This is because that $\Psi_{imag}$ is originated from the interaction phase of the NEIs and the interaction phase is expected to be zero in average for the charged liquid at equilibrium. We also present $\Psi_{var}$ in Fig.2(c). Since the BCs at the two ends of the system are fixed, the fluctuations of $\Psi_{real}$ occur in the bulk domain. The increase of $h$ excludes more ions and blocks the ions penetrating each other. In this way, the space for the free motion of the ions becomes small and fluctuation of $\Psi_{real}$ is decreased as shown in Fig.2(c). \\

We denote $\rho_k=z_k w_k e^{-z_k\Psi+\Gamma}$ for the $k$-th ionic species with $w_k$ defined by Eq.(\ref{wk}). $\rho_k$ is a complex number. The charge density of the $k$-th ionic species is the real part of $\rho_k$. In Fig.3(a), we present the charge density of the negative charges, which is denoted by $\rho_{real}$.
\begin{figure}[tbh!]
\includegraphics[scale=0.25]{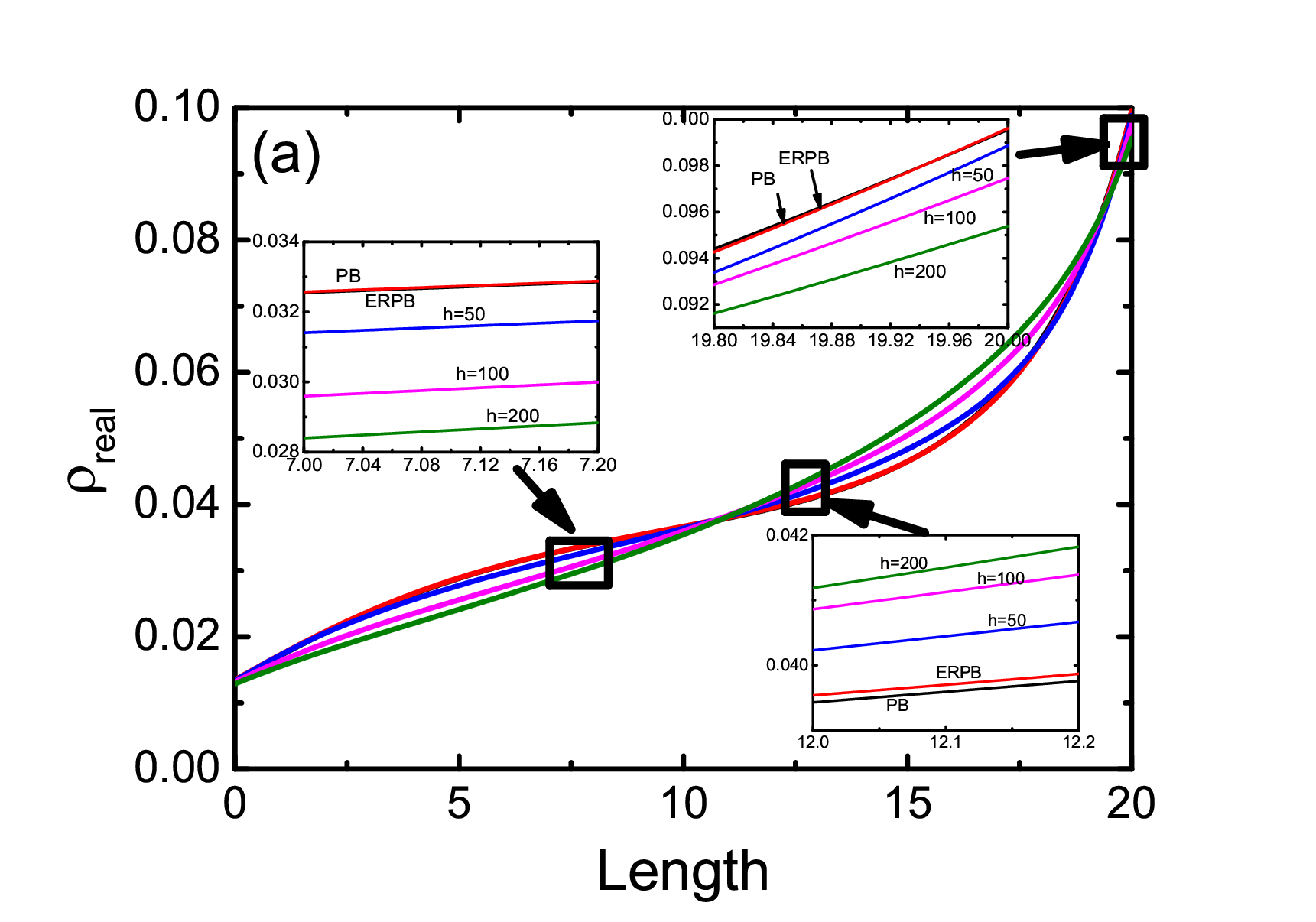}
\includegraphics[scale=0.25]{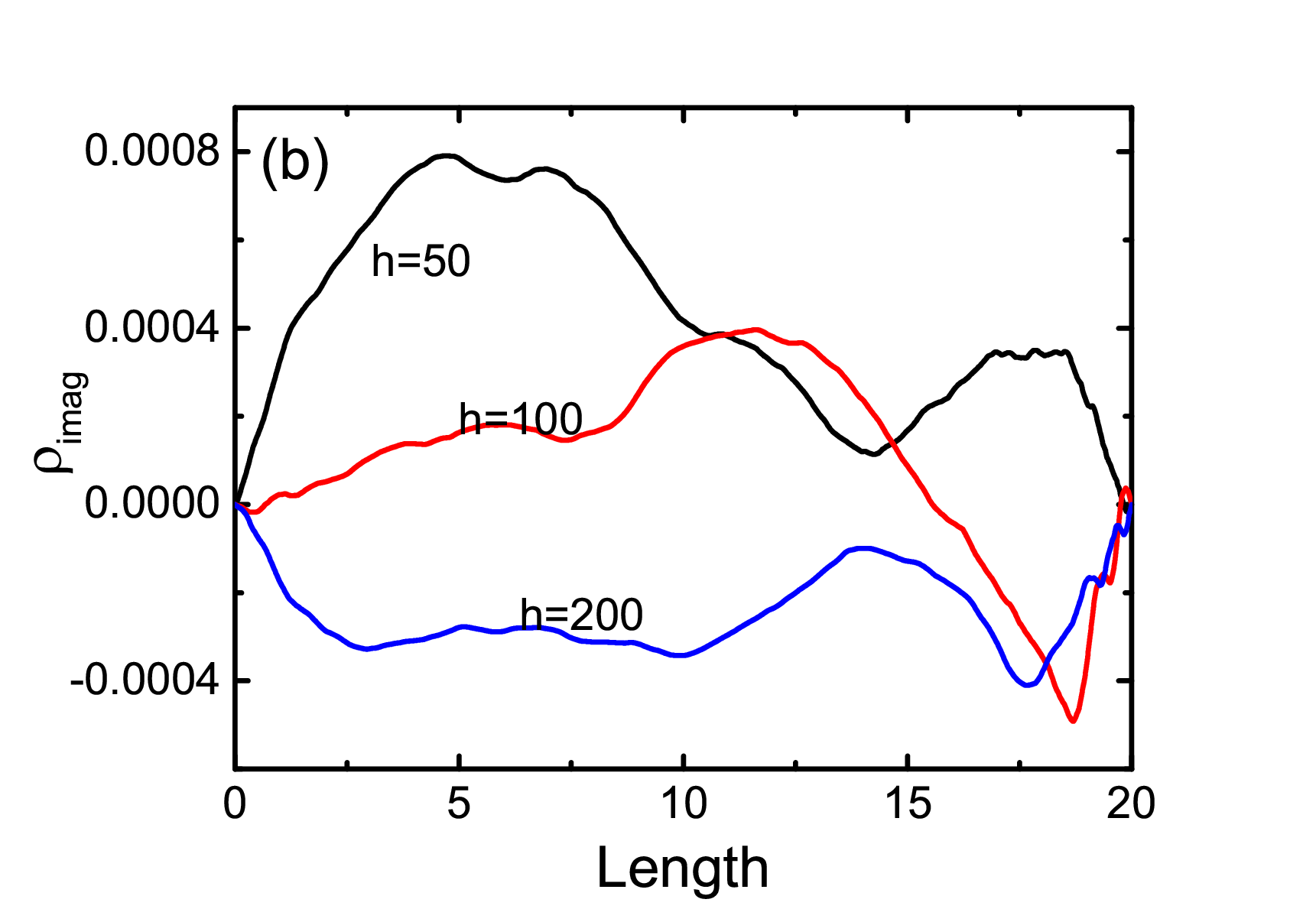}
\caption{$\rho$ for negative charges.(a) The real part of $\rho$ is modified by the steric effect. (b) The imaginary part of $\rho$ is close to zero.}
\end{figure}
The potential drop between the BD1 and the BD2 is positive, which accumulates the negative charges at the BD1 for the screening. We have enlarged three scales of the results of Fig.3(a) in the insert figures. It indicates in the insert figures that the charge densities of PB and ERPB still almost overlap, meaning that the ion fluctuation by the EI takes less effect on the ion distribution in the charged liquid where the net charge is neutral. In the insert figure close to the BD1, it is observed that the charge density decreases with the increase of $h$ for the NEPB results, showing that the steric effect excludes the negative ions and decreases the ion density at the BD1. The stronger steric effect, the smaller charge density at the BD1.\\

The negative ions excluded from the BD1 stay in the area close to the BD1 and increase the charge density in the area. We have shown such enrichment phenomena of the negative charges in the insert figure of the scale around $r=12$. In the insert figure, the charge density with $h=200$ is the largest because the most negative ions are dispersed from the BD1 due to the strongest steric effect. Due to the excluding of the negative ions at the BD1 by the steric effect, the screening of the potential at the BD1 is weakened. Therefore, more negative ions are needed to strengthen the weakened screening, which can be realized by the negative ions moving from the area close to the BD2 to the area close the BD1. Such motion of the negative ions exhausts the charge density in the area close to the BD2, which has been reflected by the insert figure of the location around $r=7$. In the insert figure of $r=7$, the charge density with $h=200$ is the lowest due to the most negative ions leaving the area close to the BD2 to the area close to the BD1 for the screening.\\

The imaginary part of $\rho$ that is denoted by $\rho_{imag}$ for the negative charges is illustrated in Fig.3(b). It shows that $\rho_{imag}$ is very small and can be neglected. It has been understood that $\rho_{imag}$ is responsible for the interaction phase of the NEIs and is expected to be zero in average. It can be concluded from Fig.3 that the steric effect can be clearly caught by Eq.(\ref{finalequ}). The conclusion from the results of the positive charges is the same to that of the negative ones, and is not shown repetitively.\\

\subsection{intersection of screening domains}
It is interesting to study the effects of the NEIs when the Debye screening domains(DSD) are intersected. The intersection of the DSD can be realized by two methods. One method is to enlarge the length of the DSD by decreasing the number density $M$ of the ions. The other method is to fix the number density to keep the length of the DSD as the constant, but decrease the length of the system. In this study, we use the latter method and keep $M=0.04$ for the charged liquid. The length of DSD is calculated to be $5$ for $M=0.04$. The DSD will intersect when the length of the liquid is smaller than $10$. For the study, the potential energy $h=100$ is used.\\

In Fig.4(a), $\Psi_{real}$ for various lengths of the liquid has been plotted. For the plot, the lengths of the liquid have been scaled to be the same for the comparison. The physical length $L$ of the system has been indicated in the plot. It shows that with the decrease of the length $L$, the electrostatic field $\Psi_{real}$ approaches to being linear. Especially for the liquid with $L=5$ where the DSD have been intersected, the result of $\Psi_{real}$ is almost a straight line. This is because the co-ions and the counter-ions are mixed to be uniform in the intersected DSD. However, such phenomena are not attributed to the NEIs only, but also can be found in the ERPB results. We present the ERPB results in the insert figure in Fig.4(a), showing the similar behaviors of $\Psi_{real}$.\\

In order to show the steric effect on $\Psi_{real}$ in the intersected DSD, we subtract the results of Fig.4(a) by the ERPB results, and present the differences in Fig.4(b). The steric effect excludes the negative ions from the BD1 to the area close to the BD1, and then increase the electrostatic potential in the area. Similarly, the steric effect decreases the electrostatic potential in the area close to the BD2 by excluding the positive ions from the BD2 to the area close to the BD2. It is also found in Fig.4(b) that the steric effect takes less influence on the electrostatic potential in the liquid with a smaller length, such as in the liquid with $L=5$. This result can be understood as the following. In the liquid with a large length, say $L=20$ where the DSD at the both boundaries do not intersect, the steric effect excludes the ions in the DSD and extends the length of the DSD in the system. In this way, the electrostatic potential is modified. However, in the system with a small length, say $L=5$ where the DSDs are intersected, the DSD have extended to the whole system already and can not be extended any more by the steric effect. Therefore, the ion distribution in the system with $L=5$ is not influenced by the steric effect.\\

The variance of $\Psi_{real}$ has also been plotted in Fig.4(c). There are two mechanisms contributing to the variance of the electrostatic field. One mechanism is from the ion fluctuation of the EI by the random field $\eta$ in Eq.(\ref{finalequ}). The variance of $\Psi_{real}$ contributed by this mechanism can be solved out from the ERPB. The other mechanism is from the steric effect by the random field $\Gamma$ in our equation. In order to get the variance contributed by the steric effect only, we subtract the variance $\Psi_{var}$ solved from the NEPB of Eq.(\ref{finalequ}) by the variance $\Psi_{var}$ solved from the ERPB. Finally, we plot the difference $\Psi_{var}$ in Fig.4(c). It shows that in the charged liquid with a smaller length, the steric effect takes less influence on the distribution of the ions. This is because in the liquid with a smaller length, there is a smaller space for the free motions of the ions and the fluctuations are suppressed. \\
\begin{figure}[tbh!]
\includegraphics[scale=0.25]{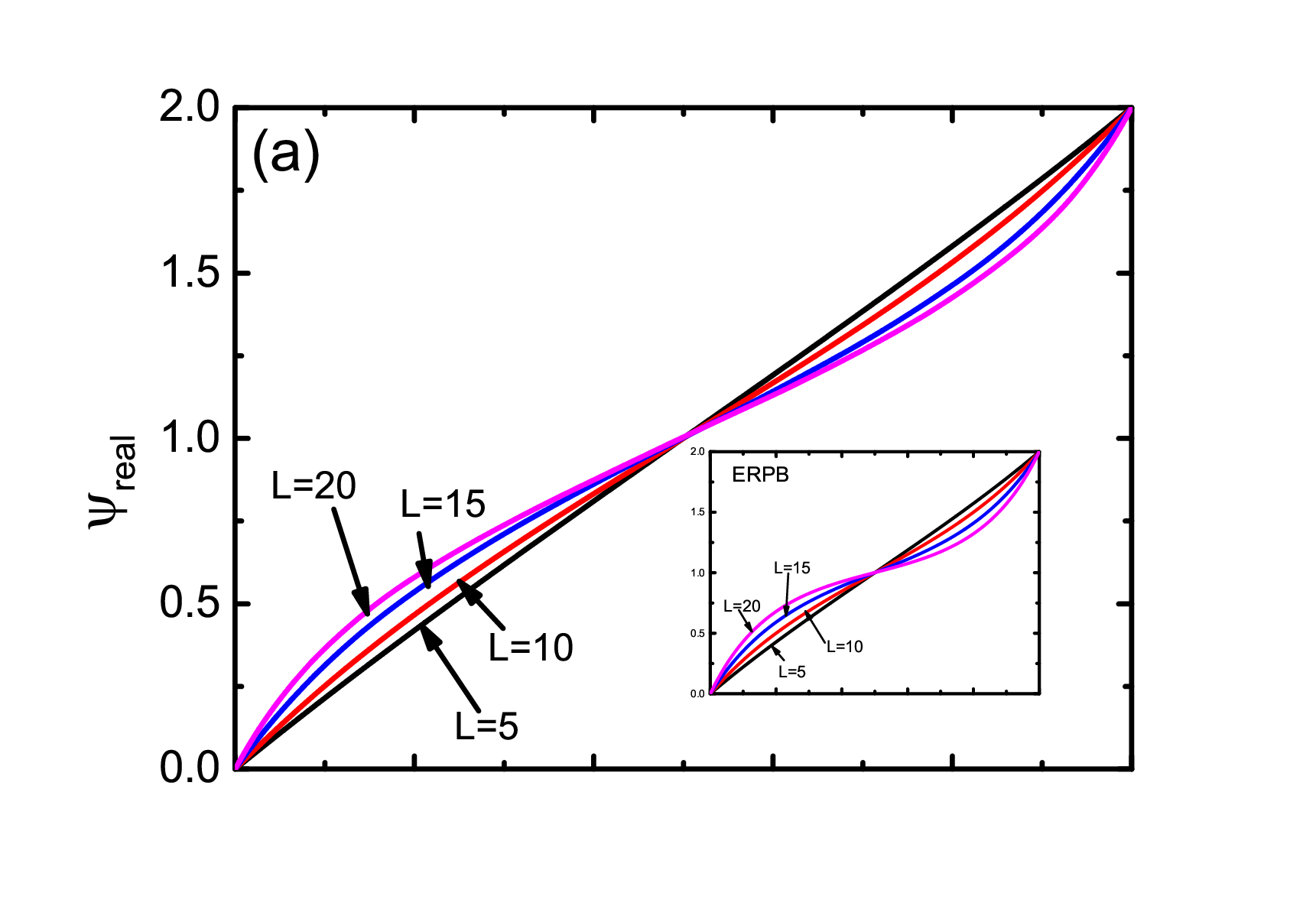}
\includegraphics[scale=0.25]{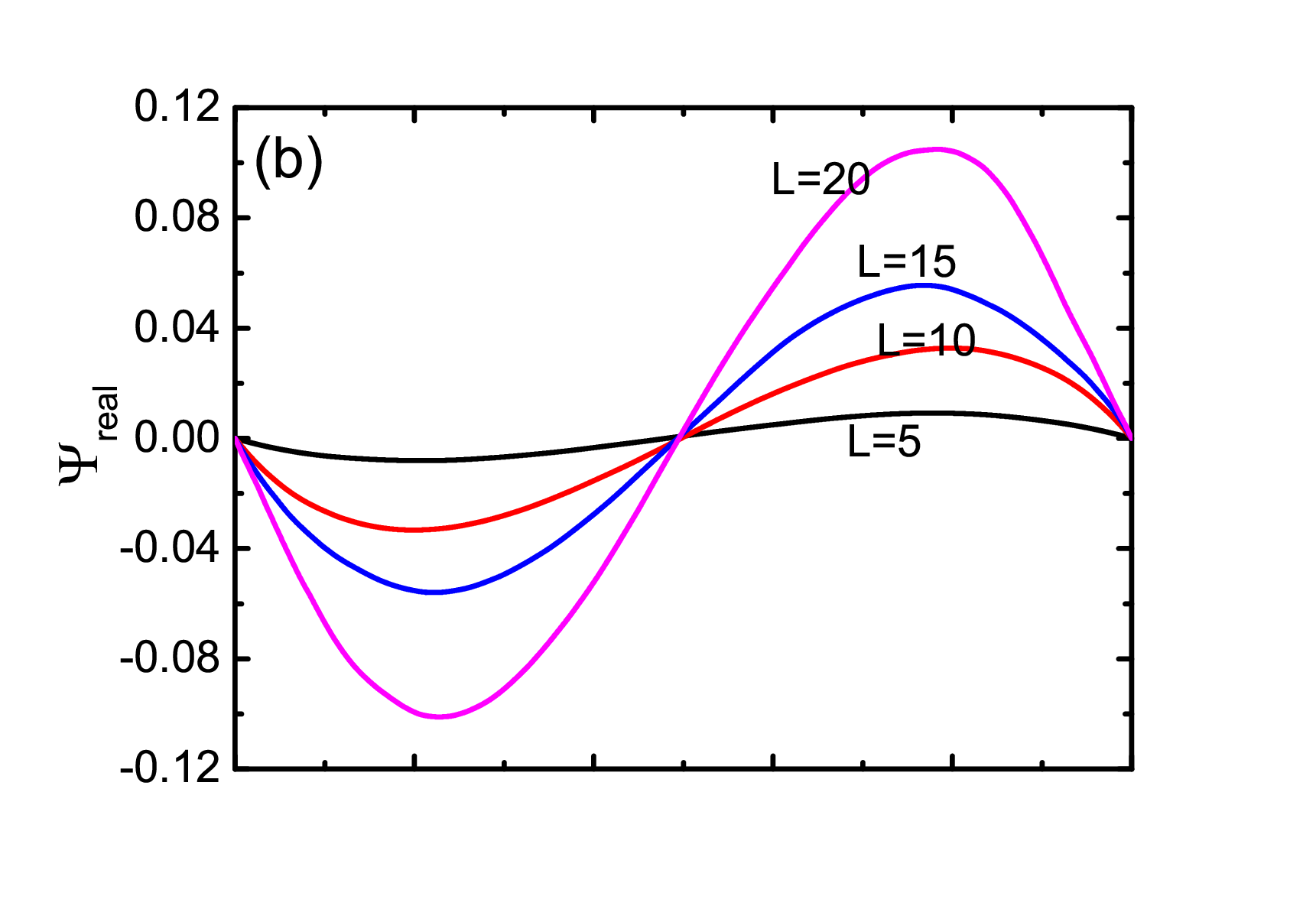}
\includegraphics[scale=0.25]{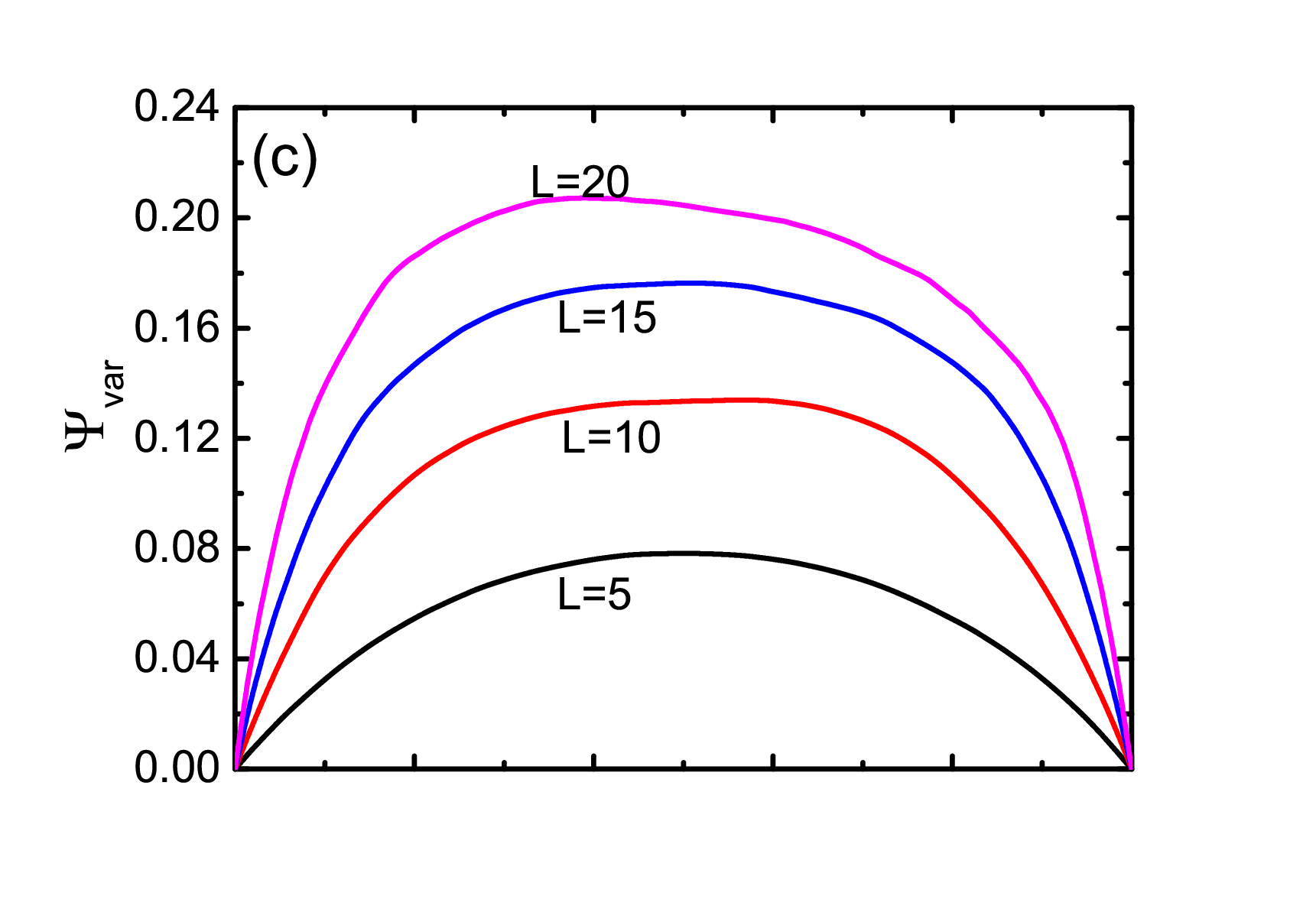}
\caption{Intersection of the Debye screening domains. (a) $\Psi_{real}$ is dependent on the length of the liquid.(b) The variation of $\Psi_{real}$ is attributed to the steric effect only. (c) The variance of $\Psi_{real}$ is induced by the steric effect only. }
\end{figure}

The steric effect in the intersected domains can also be studied by the ion density, shown in Fig.5. We present the ion density of the negative charges in fig.5(a) for various lengths of the liquid. The lengths have been scaled to be the same for the comparison. With the length of the liquid decreased, the potential drop applied on the boundaries increases the electric field in the computational domain and attracts more negative ions to the BD1 for the liquid with a smaller length. Such phenomena have also been found in the ERPB results shown in the insert figure in Fig.5(a). In order to show the steric effect only, we subtract the results of Fig.5(a) by the ERPB results, and present the differences in Fig.5(b). The positive value means the steric effect enriches the ions and the negative value means the exhaustion. The reason for the enrichment and the exhaustion has been reveal in Fig.3. It indicates that the steric effect takes less influence on the ion distribution in the liquid with a smaller length. We have enlarged the ion density close the BD1 in the insert figure in Fig.5(b), showing that the steric effect is less effective in the system with a smaller length due to the short of the space for the free motions of the ions. The results obtained for the positive charges are the same to those of the negative charges, and are not repeated here.\\
\begin{figure}[tbh!]
\includegraphics[scale=0.25]{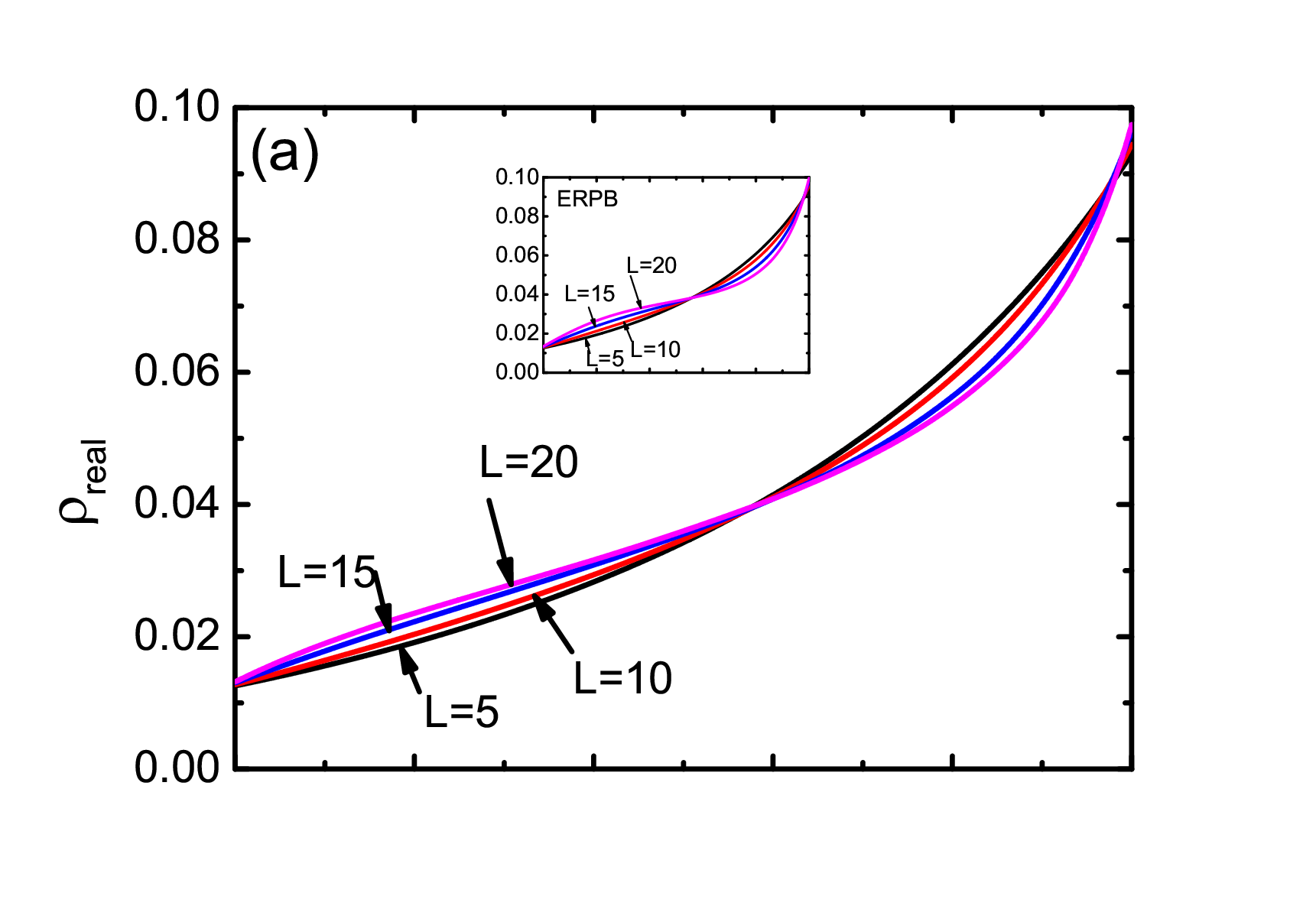}
\includegraphics[scale=0.25]{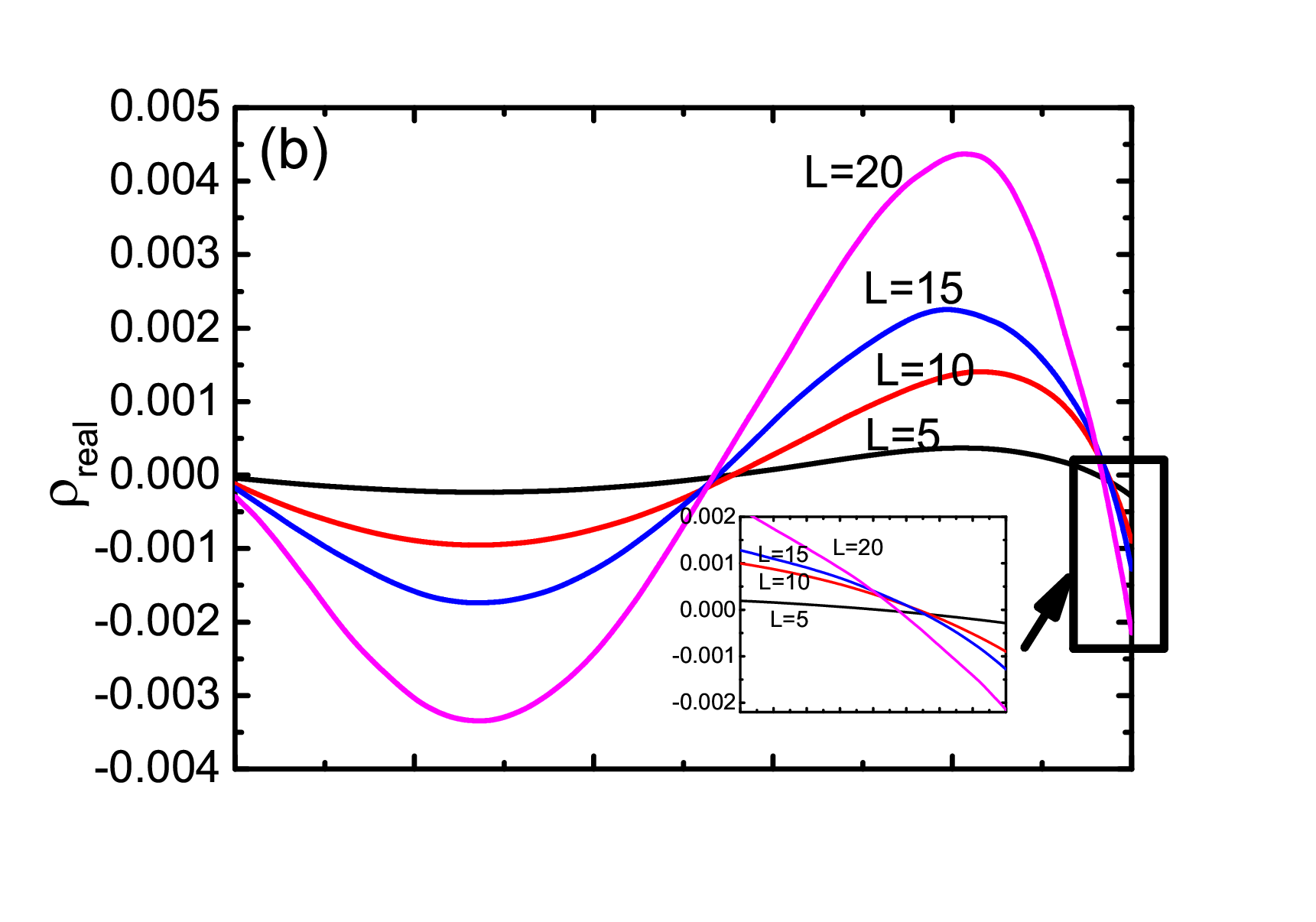}
\caption{Charge density of the negative charges in the charged liquid with various lengths.(a) The variation of the charge density is dependent on the length of the liquid. (b) The variation of the charge density is attributed to the steric effect only.}
\end{figure}

\section{conclusion}
We have derived a complex version of the PB equation for a charged liquid. In this equation, the NEIs between the ions are transformed into random fields. It is convenient to generate random numbers for the NEIs rather than calculate the NEIs directly. We take the steric effect as an example to demonstrate our theory. Results show that our theory can catch the steric effect clearly. The NEIs in our theory is general, and not limited to the steric effect.\\

The steric effect excludes the ions away from the boundaries and influences the Debye screening domains. In this way, the electrostatic potentials and the charge densities of the ions are modified. In the charged liquids where the Debye screening domains are intersected, the steric effect on the charge distribution is negligible.\\

In this study, the Stern layer can not be observed. This is because the boundary condition is set for the EI only instead of the NEIs. To apply the boundary conditions for the NEIs will be considered in our future research.  

\begin{acknowledgments}
The author kindly acknowledges Prof. Ning-Hua Tong from Renmin University of China for discussions.
\end{acknowledgments}

\appendix
\section{}
\label{eqhc}
We start from the following expression 
\begin{align}
\label{appdA}
&e^{-\beta H_c} \cdot Z_c= e^{-\beta \frac{e^2}{2}\int dr dr' \rho(r')C \rho(r)}\nonumber \\
&~~~~~~~~~~~~~~~~\cdot \int [D \xi] e^{- \left\{\frac{1}{2}\int dr dr' \xi(r')C^{-1} \xi(r)\right\}}\nonumber \\
&=\int [D \xi] e^{- \left\{\frac{1}{2}\int dr dr' \xi(r')C^{-1} \xi(r)+i\int dr \rho e\sqrt{\beta} \xi\right\}}
\end{align}
by the HST. Then substituting $C^{-1}=\nabla_{r}(\epsilon \nabla_{r'} \delta(r-r'))$ into the above equation and using the integration by parts, we have
\begin{align}
&\int dr dr' \xi(r')\nabla_{r}(\epsilon \nabla_{r'} \delta(r-r')) \xi(r)\nonumber \\
&=-\int dr' dr \xi(r')\epsilon (\nabla_{r'} \delta(r-r'))\nabla_r \xi(r)\nonumber \\
&=\int dr \epsilon (\nabla_r \xi(r))(\nabla_{r}\xi(r))
\end{align}
for the exponent in Eq.(\ref{appdA}). Finally, we divide the both sides of Eq.(\ref{appdA}) by $Z_c$ to get Eq.(\ref{hc}).

\section{}
\label{generalization}
To complement our study, we generalize Eq.(\ref{finalequ}) to a charged liquid in which the NEIs are dependent on the species. The total number of the species is denote by $K$. The number density of the ions of the $k$-th ionic species is denoted by $c_k(r)=\sum_{j=1}^{N_k}\delta(r-r_j)$ with $N_k$ the total number of the ions and $r_j$ the position vector of the $j$-th ion of the $k$-th ionic species. The number density of the total ions is $p(r)=\sum_kc_k(r)$ and the charge density in the liquid is $e\rho(r)=e\sigma+e\sum_k z_kc_k$. The FTs on $c_k(r)$ and $p(r)$ lead to $g_k'=\sum_{j=1}^{N_k}e^{-iGr_j}$ and $g'=\sum_k g_k'$ respectively. The function of the NEIs between the $k$-th ionic species and the $m$-th ionic species is denoted by $D_{km}(r,r')$. Here, $r$ and $r'$ are the position vectors of the two ions of the $k$-th ionic species and the $m$-th ionic species respectively. After the FTs, we define $T_{km}'=\int D_{km}(x) e^{-iGx}dG$ as we have defined before. Note that $D_{km}(r,r')$ is translational invariant and $x=r-r'$.\\

The Hamiltonian for the NEIs is $H_N=\frac{1}{2}\sum_{k,m}\int dr dr' p_k(r)D_{km}(r,r')p_m(r')$. We express $H_N$ in the reciprocal space by using the FT to get $H_N=\frac{1}{2}\sum_{k,m}\int dG~ g_k'[T_{km}'/(2\pi)^3]g_m'^{\dagger}$. We define a vector $\hat{g'}=[\cdots, g_k',\cdots]$ and a Matrix $\widehat{T'}$. The entry at the $k$-th row and the $m$-th column in $\widehat{T'}$ is $T_{km}'/(2\pi)^3$. We rewrite $H_N$ in the matrix form as $H_N=\frac{1}{2}\int dG~\hat{g'}\cdot\widehat{T'}\cdot\hat{g'}^{\dagger}$. The matrix $\widehat{T'}$ is real and symmetric. We introduce an unitary matrix $U$ to diagonalize $\widehat{T'}$ to get $U^{\dagger}\widehat{T'}U=\widehat{T}$. We write the $j$-th eigenvalue at the $j$-th row and the $j$-th column of $\widehat{T}$ by $\widehat{T}_{jj}=\lambda_jT_j$. Here, $T_j$ is the absolute value of $\widehat{T}_{jj}$. $\lambda_j=1$ if $\widehat{T}_{jj}$ is positive, and $\lambda_j=-1$ if $\widehat{T}_{jj}$ is negative. The $j$-th entry in the vector $\hat{g}=\hat{g'}U$ is noted by $g_j=\sum_{s}g_{s}'U_{sj}$. Then the Hamiltonian is expressed by $H_N=\frac{1}{2}\sum_j\int dG~g_j \lambda_j T_j g_j^{\dagger}$. The subscript $j$ runs over the number $K$ of the species. We discrete the reciprocal space by a lattice and introduce an infinitesimal volume $\Delta$. We rewrite $H_N=\frac{1}{2}\sum_j\sum_G ~\Delta~g_j \lambda_j T_j g_j^{\dagger}$.\\

We introduce $K$ random fields $\gamma_j$ with $j$ running from $1$ to $K$. The HST is applied, leading to 
\begin{align}
&e^{-H_N}=e^{-\frac{1}{2}\sum_j\sum_G ~\Delta~g_j \lambda_j T_j g_j^{\dagger}}\nonumber\\
&=\frac{1}{Z_{\gamma}}\int [\prod_j \mathcal{D}\gamma_j^{\dagger}\mathcal{D}\gamma_j]e^{A_{\gamma}}
\end{align}
with
\begin{align}
&A_{\gamma}=-\frac{1}{2}\sum_j\sum_G(\gamma_j^{\dagger}\gamma_j)\nonumber\\
&+\frac{1}{2}\sum_j\sum_G \sqrt{-\lambda_j} \sqrt{\Delta T_j}(\gamma_j g_j+\gamma_j^{\dagger}g_j^{\dagger}),\nonumber\\
&Z_{\gamma}=\int [\prod_j \mathcal{D}\gamma_j^{\dagger}\mathcal{D}\gamma_j]e^{-\frac{1}{2}\sum_j\sum_G(\gamma_j^{\dagger}\gamma_j)}.\nonumber
\end{align}
Here, inverse temperature $\beta$ has been dropped off due to the normalization we have defined before. The physical quantity $\beta T_j/l_B^3$ is written as $T_j$ after the normalization. Then, we define $\theta_j=\gamma_j/\sqrt{\Delta}$, and express $A_{\gamma}=A_{\gamma1}+A_{\gamma2}$ with
\begin{align}
&A_{\gamma1}=-\frac{1}{2}\sum_j\int dG(\theta_j^{\dagger}\theta_j)\nonumber\\
&A_{\gamma2}=\frac{1}{2}\sum_j\int dG~ \sqrt{-\lambda_j T_j} (\theta_j g_j+\theta_j^{\dagger}g_j^{\dagger}).\nonumber
\end{align}
The partition function of the charged liquid after the generalization reads
\begin{align}
Q&=\frac{1}{Z_cZ_{\gamma}}\left[\prod_{k=1}^K\frac{1}{N_k!\lambda_k^{3N_k}}\right]\int \left[\prod_{j=1}^K\mathcal{D}\gamma_j^{\dagger}\mathcal{D}\gamma_j \right][\mathcal{D} \xi]\times\nonumber\\
&e^{A_1+A_{\gamma_1}+A_{3}}\times \left[\prod_{k=1}^K\Lambda_k^{N_k}\right]
\end{align}
with
\begin{align}
&\Lambda_{k}=\int dr~ e^{B_{k}+\Gamma_k},\nonumber\\
&B_{k}=(-i\xi+h)z_{k},\nonumber\\
&\Gamma_k=\frac{1}{2}\sum_j\int dG~ \sqrt{-\lambda_j T_j} (\theta_j e^{-iGr}U_{kj}+\theta_j^{\dagger}e^{iGr}(U^{\dagger})_{jk}).\nonumber
\end{align}
In the above equation, the index $k$ is for the ionic species and the index $j$ is for the eigenvalues of $\hat{T}$. The NEIRF noted by $\Gamma_k$ for the $k$-th ionic species is derived from $A_{\gamma2}$ substituted by
$g_j=\sum_k g_k' U_{kj}$ and $g_j^{\dagger}=\sum_k  (U^{\dagger})_{jk}(g_k')^{\dagger}$.\\

Define $\theta_j=a_j+ib_j$, and $y_{kj}=e^{-iGr}U_{kj}$. We have $(y_{kj})^{\dagger}=e^{iGr}(U_{kj})^{\dagger}=e^{iGr}(U^{\dagger})_{jk}$. Then we simplify $\Gamma_k$ to be
\begin{align}
&\Gamma_k=\frac{1}{2}\sum_j\int dG\times\nonumber\\
&~~~ \sqrt{-\lambda_j T_j} [a_j[y_{kj}+(y_{kj})^{\dagger}]+ib_j[y_{kj}-(y_{kj})^{\dagger}]].
\end{align}
It is observed that
\begin{align}
[y_{kj}+(y_{kj})^{\dagger}]^2+i^2[y_{kj}-(y_{kj})^{\dagger}]^2=4U_{kj}(U_{kj})^{\dagger}
\end{align}
meaning that $[a_j[y_{kj}+(y_{kj})^{\dagger}]+ib_j[y_{kj}-(y_{kj})^{\dagger}]]=2\sqrt{U_{kj}(U_{kj})^{\dagger}}\vartheta_j$ with $\vartheta_j$ a Gaussian random filed following the Gaussian probability density $N(0,1)$. Thus,we have
\begin{align}
\Gamma_k=\sum_j\int dG~ \sqrt{-\lambda_j T_j}\sqrt{U_{kj}(U_{kj})^{\dagger}}\cdot\vartheta_j.
\end{align}
Eq.(\ref{finalequ}) then is generalized to be
\begin{align}
\label{generalizedequ}
-\nabla^2 \Psi=\sigma+\sum_kz_kw_ke^{-z_k\Psi+\Gamma_k}+\sqrt{\sum_kz_k^2w_ke^{-z_k\Psi+\Gamma_k}}\cdot \eta
\end{align}
with
\begin{align}
w_k=\frac{M_k\int dr}{Re[\int dr~ e^{-z_k\Psi+\Gamma_k}]}.
\end{align}
for the charge conservation.


\begin{thebibliography}{100}
\bibitem{Israelachvili}
Jacob N. Israelachvili,  \textit{Intermolecular and surface forces}, ( Academic Press, Elsevier Amsterdam, 2011).

\bibitem{Langmuir}
I. Langmuir, \textit{The role of attractive and repulsive forces in the formation of tactoids, thixotropic gels, protein crystals and coacervates} J. Chem. Phys. {\bf 6}, 873 (1938).

\bibitem{Honig}
B. Honig, A. Nicholls, \textit{Classical electrostatics in biology and chemistry} Science {\bf 268}, 1144 (1995).

\bibitem{Manciu}
M. Manciu, E. Ruckenstein, \textit{Role of the hydration force in the stability of colloids at high ionic strengths} Langmuir {\bf 17}, 7061 (2001).

\bibitem{Wennerstrom}
H. Wennerstrom, E. Vallina Estrada, J. Danielsson, M. Oliveberg,
\textit{Colloidal stability of the living cell} Proc. Natl. Acad. Sci. U.S.A {\bf 117}, 10113 (2020).

\bibitem{Parsons}
D. F. Parsons, M. Bostr{\"o}m, P. L. Nostro, B. W. Ninham, \textit{Hofmeister effects: interplay of hydration, nonelectrostatic potentials, and ion size} Phys. Chem. Chem. Phys. {\bf 13}, 12352 (2011).

\bibitem{Parsegian}
V. Parsegian, T. Zemb, \textit{Hydration forces: Observations,
explanations, expectations, questions} Current opinion in colloid $\&$ interface science {\bf 16}, 618 (2011).

\bibitem{Klaassen}
Aram Klaassen, Fei Liu, Frieder Mugele, and Igor Siretanu, \textit{Correlation between Electrostatic and Hydration Forces on Silica and Gibbsite Surfaces: An Atomic Force Microscopy Study} Langmuir {\bf 38}, 914 (2022).

\bibitem{Fedorov}
M. V. Fedorov and A. A. Kornyshev, \textit{Ionic liquids at electrified interfaces} Chem. Rev., {\bf 114}, 2978 (2014).

\bibitem{Debye}
P. Debye and E. H{\"uckel}, \textit{The theory of electrolytes. I. Lowering of freezing point and related phenomena} Phys. Z, {\bf 24}, 185 (1923).

\bibitem{Gebbie}
M. A. Gebbie, M. Valtiner, X. Banquy, E. T. Fox, W. A. Henderson, and J. N. Israelachvili, \textit{Ionic liquids behave as dilute electrolyte solutions} Proc. Natl. Acad. Sci. U.S.A, {\bf 110}, 9674 (2013).

\bibitem{Smith}
A. M. Smith, A. A. Lee, and S. Perkin, \textit{The electrostatic screening length in concentrated electrolytes increases
with concentration} J. Phys. Chem. Lett., {\bf 7}, 2157 (2016).

\bibitem{Lee}
A. A. Lee and C. S. Perez-Martinez and A. M. Smith and S. Perkin, \textit{Underscreening in concentrated electrolytes} Faraday Discuss., {\bf 199}, 239 (2017).

\bibitem{Coupette}
Fabian Coupette, Alpha A. Lee, and Andreas H{\"a}rtel, \textit{Screening lengths in Ionic Fluids} Phys. Rev. Lett., {\bf 121}, 075501 (2018).

\bibitem{Bazant}
Martin Z. Bazant, Brian D. Storey, and Alexei A. Kornyshev, \textit{Double layer in Ionic Liquids: Overscreening versus Crowding} Phys. Rev. Lett., {\bf 106}, 046102 (2011).

\bibitem{Blum}
L. Blum,  \textit{Theoretical Chemistry: Advances and Perspectives}, (Academic Press, New York, 1980).

\bibitem{Jeanmairet}
Guillaume Jeanmairet, Benjamin Rotenberg, and Mathieu Salanne, \textit{Microscopic Simulations of Electrochemical Double-Layer Capacitors} Chem. Rev., {\bf 122}, 10860 (2022).

\bibitem{Kondrat}
Svyatoslav Kondrat, Guan Feng, Fernando Bresme, Michael Urbakh, \textit{Theory and Simulations of Ionic Liquids in Nanoconfinement} Chem. Rev., {\bf 123}, 6668 (2023).

\bibitem{Zeman}
Johannes Zeman, Svyatoslav Kondrat and Christian Holm, \textit{Bulk ionic screening lengths from extremely large-scale molecylar dynamics simulations} Chem. Commun., {\bf 56}, 15635 (2020).

\bibitem{Joensson}
B. J{\"o}nsson, H. Wennerstr{\"o}m, B. Halle, \textit{Ion distributions in lamellar liquid crystals. A comparison between results from Monte Carlo simulations and solutions of the Poisson-Boltzmann equation}, J. Phys. Chem. {\bf 84}, 2179 (1980).

\bibitem{Guldbrand}
L. Guldbrand, B. J{\"o}nsson, H. Wennerstr{\"o}m, P. Linse, \textit{Electrical double layer forces. A Monte Carlo study}, J.
Chem. Phys. {\bf 80}, 2221 (1984).

\bibitem{Borukhov}
Itamar Borukhov and David Andelman, Henri Orland, \textit{Steric Effects in Electrolytes: A Modified Poisson-Boltzmann Equation}, Phys. Rev. Lett. {\bf 79}, 435 (1997).

\bibitem{Lue}
L. Lue, N. Zoeller, and D. Blankschtein, \textit{Incorppration of Nonelectrostatic Interactions in the Poisson-Boltzmann Equation}, Langmuir {\bf 15}, 3726 (1999).

\bibitem{Maggs}
A. C. Maggs and R. Podgornik, \textit{General theory of asymmetric steric interactions in electrostatic double layers}, Soft Matter {\bf 12}, 1219 (2016).

\bibitem{Loubet}
Bastien Loubet, Manoel manghi, and John Palmeri, \textit{A variational approach to the liquid-vapor phase transition for hardcore ions in the bulk and in nanopores}, J. Chem. Phys. {\bf 145}, 044107 (2016).

\bibitem{1Pelta}
J. Pelta, F. Livolant, and J.-L. Sikorav, \textit{DNA aggregation induced by polyamines and cobalthexamine},  J. Biol. Chem. {\bf 271},5656 (1996).

\bibitem{1Yoshikawa}
K. Yoshikawa,\textit{Controlling the higher-order structure of giant DNA molecules},  Adv. Drug Deliv. Rev. {\bf 52},235 (2001).

\bibitem{1Takahashi}
M. Takahashi, K. Yoshikawa, V. V. Vasilevskaya, and A. R. Khokhlov, \textit{Discrete Coil-Globule transition of single duplex DNAs induced by polyamines}, J. Phys. Chem. B {\bf 101}, 9396 (1997).

\bibitem{1Lau}
A. W. Lau, D. Lukatsky, P. Pincus, and S. A. Safran, \textit{
Charge fluctuations and counterion condensation},Phys. Rev. E {\bf 65}, 051502 (2002).

\bibitem{1Linse}
P. Linse, \textit{Mean force between like-charged macroions at high electrostatic coupling}, J. Phys.: Condens. Matter {\bf 14}, 13449 (2002).

\bibitem{1Moreira}
A. Moreira and R. Netz, \textit{Binding of similarly charged plates with counterions only}, Phys. Rev. Lett. {\bf 87}, 078301 (2001).

\bibitem{2Podgornik1}
Rudi Podgornik, Bostjan Zeks, \textit{Inhomogeneous coulomb fluid A functional integral approach}, J. Chem. Soc. Faraday Trans. 2 {\bf 84}, 611 (1988).

\bibitem{2Podgornik2}
Rudi Podgornik, \textit{An analytic treatment of the first-order correction to the Poisson-Boltzmann interaction free energy in the case of counterion-only Coulomb fluid}, J. Phys. A: Math. Gen. {\bf 23}, 275 (1990).

\bibitem{2Naji}
A. Naji and R. R. Netz, \textit{Scaling and universality in the counterion-condensation transition at charged cylinders}, Phys. Rev. E {\bf 73}, 056105 (2006).

\bibitem{2Netz1}
R. Netz and H. Orland, \textit{Beyond Poisson-Boltzmann: Fluctuation effects and correlation functions}, Eur. Phys. J. E {\bf 1}, 203 (2000).

\bibitem{2Netz2}
R. Netz and H. Orland, \textit{Variational charge renormalization in charged systems},  Eur. Phys. J. E {\bf 11}, 301 (2003).

\bibitem{2Netz3}
R. Netz, 	
\textit{Electrostatistics of counter-ions at and between planar charged walls: From Poisson-Boltzmann to the strong-coupling theory}, Eur. Phys. J. E {\bf 5}, 557 (2001).

\bibitem{2Buyukdagli1}
S. Buyukdagli, C. Achim, and T. Ala-Nissila, \textit{Electrostatic correlations in inhomogeneous charged fluids beyond loop expansion}, J. Chem. Phys. {\bf 137}, 104902 (2012).

\bibitem{2Buyukdagli2}
S. Buyukdagli, M. Manghi, and J. Palmeri, \textit{Variational approach for electrolyte solutions: From dielectric interfaces to charged nanopores}, Phys. Rev. E {\bf 81}, 041601 (2010).

\bibitem{2Buyukdagli3}
S. Buyukdagli and R. Blossey, \textit{Beyond Poisson-Boltzmann: fluctuations and fluid structure in a self-consistent theory}, J. Phys.:Condens. Matter {\bf 28}, 343001 (2016).

\bibitem{2Moreira}
A. G. Moreira and R. R. Netz, \textit{Strong-coupling theory for counter-ion distributions}, Europhys. Lett. {\bf 52}, 705 (2000).

\bibitem{2Naji1}
A. Naji, M. Kanduc, J. Forsman, and R. Podgornik, \textit{Perspective: Coulomb fluids-Weak coupling, strong coupling, in between and beyond}, J. Chem. Phys. {\bf 139}, 150901 (2013).

\bibitem{Wan1}
Li Wan, and Ning-Hua Tong, \textit{Poisson-Boltzmann equation with a random field for charged fluids}, J. Phys.: Condens. Matter {\bf 31}, 375101 (2019).

\bibitem{Wan2}
Li Wan, Shixin Xu, Maijia Liao, Chun Liu, and Ping Sheng, \textit{Self-consistent approach to global charge neutrality
in electrokinetics: A surface potential trap model}, Phys. Rev. X {\bf 4}, 011042 (2014).

\bibitem{Liu}
C.-C. Lee, H. Lee, Y. K. Hyon, T.-C. Lin, and C. Liu, \textit{New
Poisson–Boltzmann type equations: one-dimensional
solutions}, Nonlinearity {\bf 24}, 431 (2011).

\bibitem{5Camara}
R. P. Camara, G. Papastavrou, S. H. Behrens, and M.
Borkovec, \textit{Interaction between charged surfaces on the
Poisson-Boltzmann level: The constant regulation
approximation}, J. Phys. Chem. B {\bf 108}, 19467 (2004).

\bibitem{5Carnie}
S. L. Carnie and D. Y. C. Chan, \textit{Interaction free energy
between plates with charge regulation: A linearized
model}, J. Colloid Interface Sci. {\bf 161},260 (1993).

\bibitem{5Behrens}
S. H. Behrens and D. G. Grier, \textit{The charge of glass and
silica surfaces}, J. Chem. Phys. {\bf 115}, 6716 (2001).

\bibitem{5Gentil}
C. Gentil, D. C$\hat{o}$te, and U. Bockelmann, \textit{Transistor based study of the electrolyte/SiO2 interface}, Phys. Status Solidi
(a) {\bf 203}, 3412 (2006).

\bibitem{Freefem}
F. Hecht, \textit{New development in FreeFem++}, Journal of numerical mathematics {\bf 20}, 251 (2012).
\end{thebibliography}
\end{document}